\documentclass[preprint]{mn2e}

\usepackage{epsfig}

\newcommand{\Msun}{\mbox{\,M$_{\odot}$}}

\newcommand{\Msyr}{\Msun\,\mbox{yr$^{-1}$}}

\title[Evolved Cataclysmic Variables and the Progenitors of AM CVn Stars]
{Cataclysmic Variables with Evolved Secondaries and the Progenitors of
AM CVn Stars}

\author[Podsiadlowski, Han {\rm\&} Rappaport]
{Ph.~Podsiadlowski$^1$\thanks{E-mail: podsi@astro.ox.ac.uk}, Z.~Han$^2$
\& S.~Rappaport$^3$\\
\it
$^1$ University of Oxford, Nuclear and Astrophysics Laboratory, Oxford,
OX1 3RH\\
$^2$ Yunnan Observatory, National Astronomical Observatories, 
the Chinese Academy of Sciences, Kunming, 650011, China\\
$^3$ Department of Physics and Center for Space Research, Massachusetts
Institute of Technology, Cambridge, MA 02139, USA
}

\date{\today}

\volume{000}

\pagerange{\pageref{firstpage}--16} \pubyear{2001}

\begin{document}
\maketitle

\label{firstpage}

\begin{abstract}

We present the results of a systematic study of cataclysmic variables (CVs)
and related systems, combining detailed binary-population synthesis (BPS)
models with a grid of 120 binary evolution sequences calculated with a
Henyey-type stellar evolution code. In these sequences, we used 3
masses for the white dwarf (0.6, 0.8, 1.0\Msun) and seven masses for
the donor star in the range of $0.6-1.4\Msun$. The shortest orbital
periods were chosen to have initially unevolved secondaries, and the
longest orbital period for each secondary mass was taken to be just
longer than the bifurcation period ($16-22\,$hr), beyond which systems
evolve towards long orbital periods. These calculations show that
systems which start with evolved secondaries near the end or just
after their main-sequence phase become ultra-compact systems with
periods as short as $\sim 7\,$min. These systems are excellent
candidates for AM CVn stars. Using a standard BPS code, we show how
the properties of CVs at the beginning of mass transfer depend on the
efficiency for common-envelope (CE) ejection and the efficiency of
magnetic braking. In our standard model, where CE ejection is efficient,
some 10 per cent of all CVs have initially evolved
secondaries (with a central hydrogen abundance $X_c < 0.4$) and
ultimately become ultra-compact systems (implying a Galactic birthrate
for AM CVn-like stars of $\sim 10^{-3}\,$yr$^{-1}$). While these
systems do not experience a period gap between 2 and 3\,hr, their
presence in the gap does not destroy its distinct appearance. Almost
all CVs with orbital periods longer than $\sim 5\,$hr are found to
have initially evolved or relatively massive secondaries. We show that
their distribution of effective temperatures is in good agreement with
the distribution of spectral types obtained by Beuermann et al.\
(1998).

\end{abstract}

\begin{keywords}
binaries: close -- stars: AM CVn stars -- stars: cataclysmic variables
-- stars: white dwarfs -- gravitation
\end{keywords}

\section{Introduction}
AM Canum Venaticorum (AM CVn) stars are ultra-compact binaries with
orbital periods as short as $\sim 17$\,min (see Smak 1967 and Warner
\& Robinson 1972 for the original identification). Because of the
short orbital period, at least one of the binary components, and
possibly both, is believed to be a degenerate dwarf (Paczy\'nksi 1967;
Faulkner, Flannery \& Warner 1972).  Because their evolution is driven
by the loss of orbital angular momentum due to gravitational radiation,
they have also been considered important sources of background
gravitational radiation (Evans, Iben \& Smarr 1987; Hils, Bender \&
Webbink 1990; Han \& Webbink 1999).

%
%
% Table 1
%
\begin{table*}
\caption{Orbital periods and model parameters for AM CVn stars}
\tabcolsep=4pt
\begin{tabular}{lrcccccccccc}
\hline
\hline
\noalign{\vspace{1pt}}
&\multicolumn{3}{c}{Observed Parameters}&&\multicolumn{4}
{c}{Model Parameters}&&\multicolumn{2}{c}{Others}\\
\noalign{\vspace{2pt}}
\cline{2-4}\cline{6-9}\cline{11-12}
\noalign{\vspace{2pt}}
Name&\multicolumn{1}{c}{Period}
&$m_V$&Ref.&&$M_2$&$\log \dot{M}$&$\log \left(P/|\dot{P}|\right)$&
$X$&&$M_2$ (DD)&$M_2$ (He)\\
\noalign{\vspace{1pt}}
&\multicolumn{1}{c}{(s)}
&(mag)&&&(\Msun)&(\Msun\,yr$^{-1}$)&(yr)&&&(\Msun)&(\Msun)\\
\hline
AM CVn&1029&14.1\,--\,1.42&1&&
      0.077$^{+0.010}_{-0.002}$&
       -8.8$^{+ 0.1}_{- 0.0}$&
      7.9$^{+0.3}_{-0.4}$&
      0.03$^{+0.06}_{-0.00}$&
 &0.033&0.114\\
\noalign{\vspace{2pt}}
&&&&&
      0.065$^{+0.009}_{-0.002}$&
       -8.7$^{+ 0.1}_{- 0.0}$&
      8.1$^{+0.4}_{-0.3}$&
      0.01$^{+0.02}_{-0.01}$&
\\
\noalign{\vspace{2pt}}
HP Lib&1119&13.6&2&&
      0.081$^{+0.006}_{-0.003}$&
       -8.9$^{+ 0.0}_{- 0.0}$&
      7.9$^{+0.1}_{-0.1}$&
      0.11$^{+0.01}_{-0.01}$&
  &0.030&0.099\\
\noalign{\vspace{2pt}}
&&&&&
      0.061$^{+0.006}_{-0.003}$&
       -8.9$^{+ 0.1}_{- 0.0}$&
      8.2$^{+0.5}_{-0.4}$&
      0.01$^{+0.01}_{-0.01}$&
\\
\noalign{\vspace{2pt}}
CR Boo&1471&13.0\,--\,18.0&3&&
      0.096$^{+0.001}_{-0.002}$&
       -9.3$^{+ 0.0}_{- 0.0}$&
      7.7$^{+0.1}_{-0.0}$&
      0.19$^{+0.01}_{-0.00}$&
  &0.021&0.062\\
\noalign{\vspace{2pt}}
&&&&&
      0.047$^{+0.004}_{-0.003}$&
       -9.5$^{+ 0.1}_{- 0.1}$&
      8.6$^{+0.5}_{-0.5}$&
      0.02$^{+0.01}_{-0.02}$&
\\
\noalign{\vspace{2pt}}
V803 Cen&1611&13.2\,--\,17.4&2&&
      0.099$^{+0.003}_{-0.001}$&
       -9.3$^{+ 0.0}_{- 0.0}$&
      7.7$^{+0.0}_{-0.1}$&
      0.21$^{+0.00}_{-0.00}$&
  &0.019&0.054\\
\noalign{\vspace{2pt}}
&&&&&
      0.043$^{+0.004}_{-0.004}$&
       -9.6$^{+ 0.1}_{- 0.1}$&
      8.7$^{+0.5}_{-0.4}$&
      0.03$^{+0.01}_{-0.03}$&
\\
\noalign{\vspace{2pt}}
CP Eri&1724&16.5\,--\,19.7&2&&
      0.100$^{+0.003}_{-0.021}$&
       -9.4$^{+ 0.0}_{- 0.0}$&
      7.8$^{+0.8}_{-0.1}$&
      0.22$^{+0.00}_{-0.12}$&
 &0.017&0.048\\
\noalign{\vspace{2pt}}
&&&&&
      0.040$^{+0.005}_{-0.004}$&
       -9.8$^{+ 0.1}_{- 0.1}$&
      8.7$^{+0.5}_{-0.4}$&
      0.03$^{+0.01}_{-0.03}$&
\\
\noalign{\vspace{2pt}}
GP Com&2970&15.7\,--\,16.0&2&&
      0.037$^{+0.022}_{-0.003}$&
      -10.7$^{+ 0.3}_{- 0.1}$&
      9.9$^{+0.2}_{-0.6}$&
      0.23$^{+0.02}_{-0.07}$&
  &0.008&0.019\\
\noalign{\vspace{2pt}}
&&&&&
      0.031$^{+0.008}_{-0.009}$&
      -10.7$^{+ 0.2}_{- 0.3}$&
      9.5$^{+0.6}_{-0.4}$&
      0.20$^{+0.05}_{-0.20}$&
\\
\noalign{\vspace{2pt}}
\hline
\end{tabular}\\
\parbox{6truein}{
\noindent{Note. ---} $m_V$: visual magnitude; $M_2$: mass of secondary;
$\dot{M}$: mass-transfer rate; $ \left(P/|\dot{P}|\right)$: timescale
for orbital period change;  $X$: surface hydrogen abundance. Most
model parameters are from the present study, where the values in the
first row are for systems before the period minimum (i.e. when
their orbital period decreases) and the values in the second row are for
systems after the period minimum (i.e. when their orbital period increases).
The alternative values for $M_2$ listed under
`Others' were taken from NPVY (as well as the observational parameters and
references); the column marked `DD' assumes a double-degenerate model
and the column marked `He' assumes a semi-degenerate helium-star model.}\\

\smallskip
\parbox{6truein}{
\noindent{References. ---} (1) Patterson et al.\ (1993), (2) Warner (1995),
(3) Provencal et al.\ (1997).}
\end{table*}

At present there are two popular models for the formation of AM CVn
stars: (1) a double-degenerate model and (2) a model where the
secondary is a semi-degenerate helium star.  In the double-degenerate
model (Tutukov \& Yungelson 1979; Nather, Robinson \& Stover 1981), AM
CVn stars consist of two degenerate dwarfs that formed a close binary
after experiencing one or two common-envelope phases (Paczy\'nski
1976; Iben \& Tutukov 1984; Webbink 1984); the lighter component of
the system (typically a He white dwarf) is now observed in the process
of transferring mass to the more massive companion in a stable manner
(for a detailed analysis of the stability of mass transfer of double
degenerates, see Han \& Webbink 1999). In the second model, several
phases of mass transfer first lead to the formation of a detached
system consisting of a non-degenerate low-mass helium star (typically
$\sim 0.4 -0.6\Msun$) and a white-dwarf companion.  If the orbital
period of this system is short enough, angular-momentum loss due to
gravitational radiation will cause the orbit to decay until the helium
star fills its Roche lobe and starts to transfer mass (at an orbital
period $\la 1\,$hr); the system will now appear as an AM CVn star
(Savonije, de Kool \& van den Heuvel 1986; Iben \& Tutukov 1991). The
orbital period will decrease further, reaching a minimum period of
$\sim 10\,$min, and then start to increase again (while the secondary
becomes semi-degenerate).

Nelemans et al.\ (2001a; NPVY) have recently published a thorough
binary-population synthesis (BPS) study of AM CVn stars and found
that both of these evolutionary channels could be of comparable importance;
this depends, however, on many uncertain factors in the modelling.
(We refer to this paper and Warner [1995] for excellent reviews of AM
CVn stars and further references.) In Table~1 we list the observational
properties of 6 known AM CVn stars, taken directly from NPVY, and
some model parameters from NPVY and the present
investigation (\S~3)\footnote{NPVY also list RX J1914+24 as a further
AM CVn candidate. In this system, Cropper et al.\ (1998) have observed
a period of 569\,s, which they interpret as the spin period of a magnetic
white dwarf; they further suggest that the spin of the white dwarf may be
magnetically locked to the orbit, in which case this period would also
constitute the orbital period of the system, making it the
binary with the shortest known orbital period. Since we consider this
interpretation uncertain at the present time, we have not listed this
system, but note that the model described in \S~2 can explain systems
with periods as short as $\sim 7\,$min.}.

In addition to these two channels there is a third channel that leads to
the formation of ultra-compact systems, which is well established in
the context of ultra-compact systems containing a neutron star
(Tutukov et al.\ 1987; Fedorova \& Ergma 1989; Podsiadlowski,
Rappaport \& Pfahl 2001 [PRP]).  It requires only that a normal H-rich
star ($\sim 1\Msun$) starts to transfer mass near the end of or just
after hydrogen core burning.  Such systems become ultra-compact
binaries with periods as short as $\sim 10\,$min where the secondary
is initially non-degenerate and hydrogen-rich, but increasingly
becomes degenerate and helium-rich during its evolution. NPVY
dismissed this channel as one of low probability. However, in a
recent study of low-/intermediate-mass X-ray binaries (LMXBs/IMXBs),
PRP found that the initial period range that leads to ultra-compact
systems for a secondary with an initial mass of 1\Msun\ is quite
large, 13 to 18 hr, and suggested that this is the reason why
ultra-compact LMXBs are so common in globular clusters. In addition,
this alternative channel depends on relatively few uncertainties in
the theoretical modelling of the binary evolution. It is one of the
purposes of this paper to demonstrate that this channel indeed provides
a viable alternative channel for the formation of AM CVn stars.

During their early evolution, such systems will appear as cataclysmic
variables (CVs) with evolved secondaries. Recently, Beuermann et al.\
(1998) and Smith \& Dhillon (1998) have shown that, for orbital periods
longer than $\sim 5\,$hr, there is a large spread in the observed
spectral types of the secondaries in CVs at a given orbital period
and suggested
that a large fraction of these systems had companions with evolved
secondaries. These evolved systems are excellent candidates for the
progenitors of AM CVn systems. However, these observations also pose
an immediate problem for the standard explanation of the period gap
in CVs between 2 and 3 hr, the disrupted-magnetic braking model
(Rappaport, Verbunt \& Joss 1983; Spruit \& Ritter 1983). In this
model, magnetic braking dramatically decreases
when the secondary becomes fully convective (at an orbital
period of $\sim3\,$hr), allowing it to relax to
thermal equilibrium and to become detached. Mass transfer starts again
once angular-momentum loss due to gravitational radiation has brought
the secondary back into contact which occurs at an orbital period of
$\sim2\,$hr (for detailed studies see Kolb 1993; Howell, Nelson \& Rappaport
2001 [HNR]).  However, it is well established that CVs
that start with only a slightly evolved secondary experience
a gap at shorter orbital periods or no period gap at all (Pylyser \&
Savonije 1989; PRP and \S~2).  Hence if the majority of all CVs had an
evolved companion initially, this would challenge the standard
explanation for the period gap.  However, as we will show in this
contribution there is no serious conflict between the existence of
evolved systems and the period gap, even for quite standard
assumptions\footnote{Baraffe \& Kolb (2000) attempted to solve this
apparent conflict by proposing that evolved systems have much higher
mass-transfer rates than obtained in standard binary evolutionary
calculations and `bounced' before reaching the period gap, i.e. never
reach orbital periods $\la 3\,$hr. However, this constitutes a
significant departure from the standard model for the evolution of CVs
and would in our view compromise some of the fundamental assumptions of
this model, that has been so successful in explaining key observational
features of CVs (see e.g. Kolb 1983; HNR]).}.

In \S~2 of this paper we briefly summarize our binary stellar evolution
calculations and present detailed calculations to illustrate the
formation of ultra-compact, AM CVn-like systems. In \S~3 we use a
large grid of these binary evolution sequences and integrate them into
a standard binary population synthesis code to simulate the main properties
of CVs and related objects. In \S~4 we discuss
the implications of these results for ultra-compact systems (like
AM CVn's) and CVs with evolved secondaries and compare our simulations
with observations.

\section{Binary stellar evolution calculations}

All binary evolution calculations were carried out with a standard
Henyey-type stellar evolution code (Kippenhahn, Weigert \& Hofmeister 1967),
using solar metallicity ($Z=0.02$), up-to-date opacities and
a mixing-length parameter $\alpha=2$ (for a detailed description see PRP).
We include angular-momentum loss due to gravitational radiation and
magnetic braking. For the latter we use the formalism of
Rappaport, Verbunt \& Joss (1983; their eq.~36 with $\gamma=4$),
which is based on the magnetic-braking law of Verbunt \& Zwaan (1981).
We only include magnetic braking, if the secondary (the mass-losing component)
has a sizeable convective envelope, and assume that it stops abruptly
when the secondary  becomes fully convective (Rappaport et al.\ 1983;
Spruit \& Ritter 1983). We further assume that mass transfer is completely
non-conservative, i.e. that all the mass lost from the secondary is ejected
from the system (e.g. in the form of nova explosions or a disc wind), where
the matter that is lost from the system carries with it the specific orbital
angular momentum of the white dwarf.

%
%
% Figure 1
%
\begin{figure*}
\centering
\epsfig{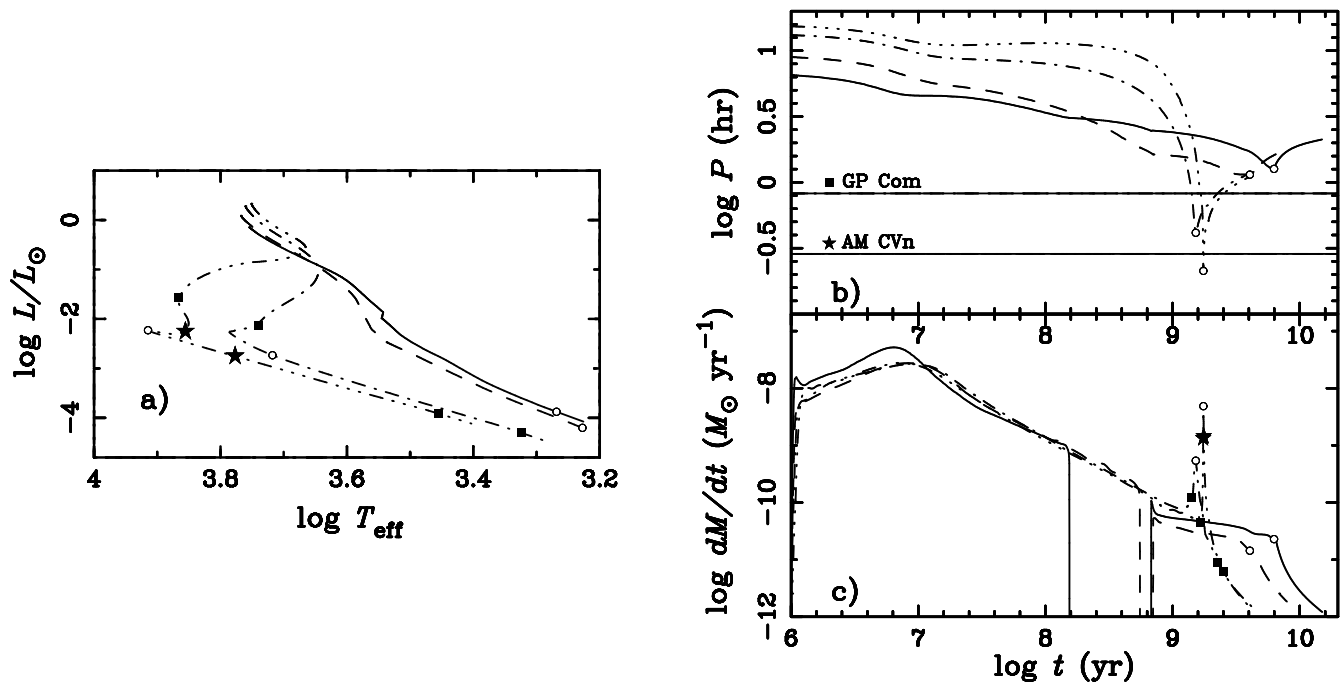}
\caption{\small Selected binary sequences illustrating the formation of AM
CVn systems. All systems initially consist of a white dwarf of
0.6\Msun\ and a normal-star companion of 1\Msun\ for different initial
orbital periods (line styles from short to long initial period: solid,
dashed, dot-dashed, dot-dot-dot-dashed). The individual panels show
the evolutionary tracks in the H-R diagram (left), the evolution of
orbital period (top right) and mass-transfer rate (bottom right) as a function
of time since the beginning of mass transfer.  The open circles show
where the systems in the individual sequences reach their period
minimums.  Other symbols indicate when the systems pass through the
periods of two representative AM CVn systems (squares: GP Com, stars:
AM CVn).
}
\end{figure*}

In Figure~1 we present the results of four binary sequences for a white-dwarf
accretor with mass $M_1=0.6\Msun$ and a normal donor star with an initial
mass $M_2=1\Msun$ for different evolutionary phases of the donor
at the beginning of mass transfer: unevolved donor (solid curves),
evolved donor with a central hydrogen abundance of $X_c=0.10$
(dashed curves), evolved donor with hydrogen-exhausted cores
of mass $M_c = 0.037\Msun$ (dot-dashed curves) and $M_c = 0.063\Msun$
(dot-dot-dot-dashed curves), respectively. (This figure is analogous
to fig.~16 in PRP.)

First consider the sequence with an initially unevolved secondary
(dashed curves). Since the secondary is initially significantly more
massive than the white-dwarf primary, it cannot remain in near
thermal equilibrium,
and mass transfer initially occurs on a thermal timescale, reaching
a maximum mass-transfer rate of $\sim 5\times 10^{-8}\Msun$\,yr$^{-1}$.
Even after the mass ratio has been reversed, it cannot relax to thermal
equilibrium since the timescale for magnetic braking, the dominant
mechanism driving mass transfer at this stage, is comparable to the
Kelvin-Helmholtz timescale of the secondary ($\sim 2\times 10^{8}\,$yr).
This changes abruptly when the secondary becomes fully convective and
magnetic braking is assumed to become ineffective (at an orbital
period of $3.07\,$hr in this example). Now it can relax and shrink to
its equilibrium radius, associated with a kink in the Hertzsprung-Russell
(H-R) diagram (left panel in Fig.~1). The system remains detached until
angular momentum loss due to  gravitational radiation brings the secondary
back into contact with its Roche lobe (at an orbital period of $2.45\,$hr).
Gravitational radiation, which is the dominant mass-transfer driving
mechanism now, continues to drive the orbital evolution of the system,
but at a lower rate than magnetic braking did, leading to a
lower mass-transfer
rate below the period gap. The orbital period decreases until it reaches
a minimum at 75$\,$min and then starts to increase again (while the
secondary is becoming increasingly degenerate).

There are several points to note about this sequence. First, the
period gap between 2.45 and 3.07\,hr in this sequence
(as well as in all other sequences with unevolved secondaries) is
somewhat smaller than the period gap found in other studies of CVs
(e.g. Kolb 1993; HNR) which were specifically designed to explain the
observed period gap between 2 and 3\,hr (Ritter \& Kolb 1998). This
is not surprising since we did not calibrate our magnetic-braking law
to reproduce the period gap. The fact that our gap is too small implies
that our model somewhat underestimates the amount of magnetic braking
just above the gap. Second, the period minimum found in our sequences
is somewhat higher (at 75\,min) than the period minimum obtained in other
recent studies of CVs ($\sim 65\,$min, Kolb \& Baraffe 1999; HNR). Since
these other investigations use more sophisticated equations of state, more
appropriate in the regime of very-low-mass secondaries near the period
minimum, their quoted values are likely to be more realistic and should
be preferred. Despite these limitations, we do not believe that these
differences will affect any of the main conclusions in this paper.

Consider now the dashed sequence, where the secondary starts to
transfer mass when it is already quite evolved, and its central
hydrogen abundance (by mass) has decreased to $X_c=0.10$. In this
case, the secondary becomes fully convective at a much lower mass
($0.14\Msun$), implying a much shorter orbital period when the system
becomes detached. Since the magnetic-braking timescale is much longer
at this shorter period, the secondary is less out of thermal
equilibrium and less oversized relative to is equilibrium radius.  As
a consequence, the system experiences only a very short period gap
between 1.30 and 1.45\,hr.  We find quite generally that for systems
to experience a significant period gap between 2 and 3\,hr, they have
to be relatively unevolved initially; typically the central hydrogen
abundance at the beginning of mass transfer has to be larger than
$\sim 0.40$ (see Table~A1). After gravitational radiation has brought
the system back into contact, the subsequent evolution is similar to
the previous case, but reaches a period minimum at a substantially lower
orbital period of 55\,min.

In the two other cases, where the secondaries already have
hydrogen-exhausted cores of $0.037\Msun$ and $0.063\Msun$
at the beginning of mass transfer, respectively, there is no period
gap, and the secondaries transform themselves into essentially pure
helium white dwarfs (with some traces of hydrogen, 5 and 4 per cent
left initially). These systems reach extremely short
orbital periods of 22 and 11\,min, respectively. Note that near the
minimum period (open circles in Fig.~1), the secondaries are still
relatively luminous and hot, but then cool rapidly and join the
appropriate white-dwarf cooling curve. Since the gravitational
radiation timescale becomes very short at these short orbital
periods, the mass-transfer rate shows a sharp spike near the period
minimum. Both of these sequences pass through the period range of many
AM CVn stars, once when the orbital period is decreasing and once
after the period minimum (very similar to the behaviour of the
analogous ultra-compact systems with neutron-star primaries; see
Fedorova \& Ergma 1989 and PRP). One possible way of distinguishing
between these two passages is that the secondaries in systems before
the period minimum have larger amounts of hydrogen left in
their envelopes. Systems that have hydrogen-exhausted cores
at the beginning of mass transfer transform themselves into essentially
pure helium white dwarfs near the period minimum
(see the column marked `$X$' in Table~1 and \S~4).  In addition,
the rates of change in the orbital period will be of opposite sign.

In order to investigate this channel for ultra-compact AM CVn-like
systems more systematically, we have carried out a large number of binary
evolution calculations, where we varied the initial masses of both
components and the evolutionary phase at the beginning of mass
transfer.  Altogether we performed 120 such calculations for three
masses of the primary ($M_1= 0.6$, 0.8 and 1.0\Msun) and seven masses
of the secondary ($M_2 = 0.6$, 0.8, 0.9, 1.0, 1.1, 1.2 and 1.4\Msun).
The shortest orbital period for each combination of masses was chosen
so that the secondary filled its Roche lobe on the zero-age main
sequence (i.e. when it was completely unevolved). The longest orbital
period was determined so that it was slightly larger than the
bifurcation period above which systems evolve towards longer orbital
periods rather than shorter orbital periods (see e.g. Pylyser \&
Savonije 1989 and PRP). The value of the bifurcation period was found
to vary from $\sim 16\,$hr for the sequences with a 1\Msun\ secondary
to $\sim22\,$hr for the most massive sequences. In cases
where the main-sequence lifetime was longer than $12\times
10^{10}\,$yr (for the lower-mass secondaries), the largest period was
chosen by the constraint that the secondary can evolve to fill its
Roche lobe in less than $12\times 10^{10}\,$yr.

Some of the key characteristics of each sequence are given in the
appendix in Table~A1. Our grid of models covers a wide range of
evolutionary behaviours, from systems that are exclusively driven by
angular-momentum loss due to magnetic braking and gravitational
radiation (classical CVs) to systems that experience rapid mass
transfer on a thermal timescale and would appear as supersoft X-ray
sources (see e.g. Rappaport, Di\,\,Stefano \& Smith 1994; Langer et al.\
2000; King et al.\ 2001). We note that in all calculated sequences,
mass transfer was dynamically stable at all times. The only sequence
that was close  to dynamical instability was the sequence with a
0.6\Msun\ white dwarf and an initially unevolved 1.4\Msun\ secondary.

\section{Binary population synthesis}

The determination of the statistical properties of CVs
and AM CVn stars requires a binary population
synthesis (BPS) study, which starts with a sample of primordial binaries
with given distributions of component masses and orbital periods, follows
the various evolutionary paths encountered and derives the statistical
properties of the systems of interest. The BPS method utilized in this
investigation differs from most previous studies of this kind in so far
as we use detailed binary evolution sequences to model the evolutionary
phase of interest, the CV phase, instead of relying on simplified
prescriptions for the evolution. This allows us to also include
phases of mass transfer
that occur on a thermal timescale and phases where the chemical composition
of the mass donor changes, which cannot be reliably treated
with simplified prescriptions (unlike the case of classical CVs;
see e.g. HNR). More specifically, we directly link the grid of 120
binary sequences described in the previous section with a standard BPS code:
we first use the BPS code to determine the distributions of masses and orbital
periods at the beginning of the CV phase and to assign relative
weights to the individual sequences, and then we use these properly
weighted sequences to construct the statistical properties of CVs.

\subsection{The binary population synthesis code}

The BPS code we use in this study was originally
developed by Han (1995) and is described in detail in Han, Podsiadlowski
\& Eggleton (1995; HPE). Since then the code has been continuously
revised and updated, as described in Han et al.\ (2001). Here we restrict
ourselves to listing some of the main assumptions as relevant for
the present investigation.

The code uses a standard Monte-Carlo technique, where we typically
simulate the evolution of $10^7$ systems. For the initial
distributions, we assume a Miller-Scalo distribution for the primary
(Miller \& Scalo 1979), take the mass-ratio distribution to be flat,
and adopt a distribution of initial orbital separation that is flat in
$\log a$. We further assume that the star-formation rate is constant
over the age of the Galaxy (taken to be $12\,$Gyr), forming
one binary system with a primary mass larger than 0.8\,\Msun\ per year
in the Galaxy (this corresponds to an integrated Galactic
star-formation rate by mass of 3.5\Msun\,yr$^{-1}$). This
star-formation rate implies the formation of $\sim 1$ white dwarf
(either single or in a binary) per year in the Galaxy.

All of these assumptions are fairly standard. Where our method differs
significantly from most other BPS studies is in the treatment of
the common-envelope and the spiral-in phase (the conditions for
the occurrence of dynamical mass transfer are described in
detail in Han et al.\ 2001).
As in most other studies, we assume that the common envelope is ejected
when the change in orbital energy, $\Delta E_{\rm orb}$, times some
factor, $\alpha_{\rm CE}$, exceeds the binding energy of the envelope.
Unlike most other studies, however, we do not use an analytic
formula for the binding energy, but use the values obtained from detailed
stellar calculations and also include a contribution from the thermal
energy of the envelope (see HPE for details and also
Dewi \& Tauris 2000). Our envelope ejection criterion is written as
\begin{equation}
\alpha_{\rm CE}\,|\Delta E_{\rm orb}| > |E_{\rm gr} + \alpha_{\rm th}
\,E_{\rm th}|,
\end{equation}
where $E_{\rm gr}$ is the gravitational potential energy of the
envelope and $E_{\rm th}$ its thermal energy, which in particular
includes the ionization energy. The inclusion of the ionization energy
can change the orbital-period distribution immediately after the
ejection of the common envelope (CE) quite dramatically, since for
evolved stars on the asymptotic-giant branch (AGB), the total energy
of the envelope becomes small and ultimately zero if the ionization
energy is included (see Han, Podsiadlowski \& Eggleton 1994). If this
ionization energy can be extracted in the CE ejection process, very
little orbital shrinking is required to provide enough energy to eject
the loosely bound envelopes on the AGB, leading to relatively long
post-CE orbital periods. To account for this possibility, we therefore use
two parameters rather than one, $\alpha_{\rm CE}$ and $\alpha_{\rm
th}$, in our definition of the CE ejection criterion. In the following, we
will use two sets of parameters: a set for efficient CE ejection (with
$\alpha_{\rm CE}=\alpha_{\rm th}=1$), and a less efficient one (with
$\alpha_{\rm CE}= 0.5$, $\alpha_{\rm th}=0$).  The first set assumes the
maximum CE-ejection efficiency consistent with energy conservation and
leads to a post-CE distribution of orbital periods extending to
$\sim 420\,$d, while the second produces a distribution that
is limited to less than $\sim 6\,$d (see e.g. figure~4 in HPE). We generally favour the
more efficient ejection criterion in order to be able to explain
certain systems containing a white dwarf (some symbiotic binaries and
barium stars; see the discussions in HPE and Han et al.\ 1995), which
have orbital periods that are relatively long ($\ga 100\,$d), but
appear to be too short to have avoided a CE phase altogether.

\subsection{The distributions at the beginning of mass transfer}

%
% Figure 2
%
\begin{figure}
\centering
\epsfig{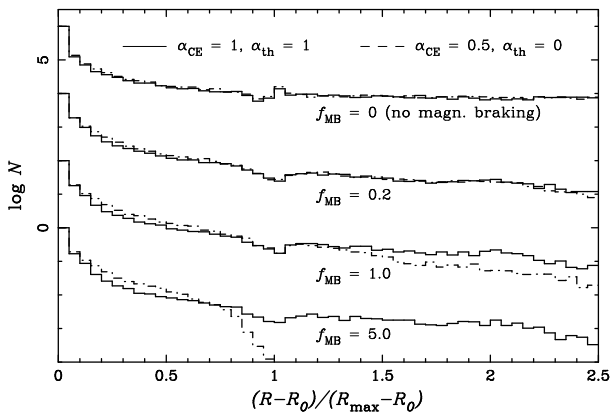}
\caption{Number distribution
of the evolution parameter $r=(R-R_{0})/(R_{\rm max} - R_{0})$ for
different BPS models and assumptions about the efficiency of magnetic
braking, where $R$ is the radius of the secondary at the
beginning of mass transfer, $R_0$ and $R_{\rm max}$ are the radii of a
star of the same mass at the beginning of hydrogen burning and at the
point of hydrogen exhaustion, respectively. Solid histograms are
for models with efficient common-envelope (CE) ejection ($\alpha_{\rm CE} = 1$
and $\alpha_{\rm th} =1$), dot-dashed histograms for models with less
efficient CE ejection ($\alpha_{\rm CE} = 0.5$ and $\alpha_{\rm th} =0$).
The histograms for different efficiencies of magnetic
braking (as indicated) have been shifted by two orders of magnitude
relative to each other for clarity.
}
\end{figure}

The progenitors of CVs are generally believed to be relatively wide
systems with orbital periods of several years that experience a CE and
spiral-in phase, leaving a tight binary with a period of less than a
few days after the envelope has been ejected (Paczy\'nski 1976).  In
most cases, the system will be detached immediately after the
CE phase, but will come back into contact due to either of two
complementary effects: the first is the orbital shrinking due to the
loss of orbital angular momentum caused by magnetic braking and
gravitational radiation; the second is the expansion of the secondary
due to it own nuclear evolution (for stars that are massive enough
that they evolve appreciably in a Hubble time). It is the balancing of
these two effects that determines the evolutionary state of the
secondary when it starts to transfer mass. As already discussed in
\S~2, the standard explanation for the period gap requires that most
secondaries have to be relatively unevolved at the beginning of mass
transfer, which implies that for the majority of stars the timescale
for angular-momentum losses has to be shorter than the evolutionary timescale
of the secondary. This also has the somewhat unfortunate consequence
that the distribution of the evolutionary states of the secondaries
at the beginning of mass transfer is rather sensitive to the assumed
strength of magnetic braking, which is still rather uncertain.

%
% Figure 3
%
\begin{figure}
\centering
\epsfig{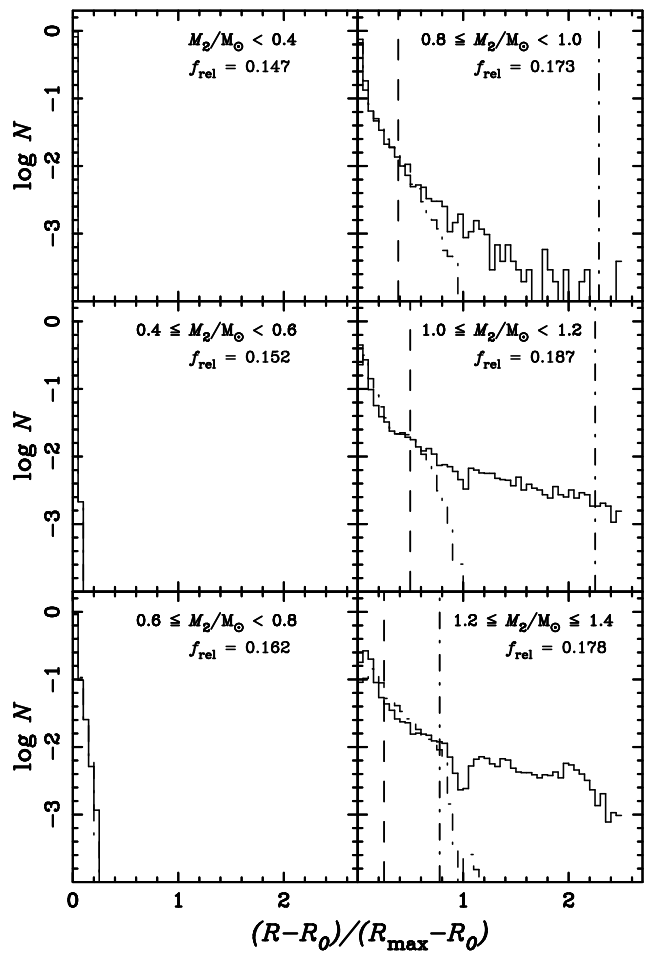}
\caption{Number distributions as a function of evolution parameter (see
Fig.~2) for CVs at the start of mass transfer for different mass ranges for the
secondary, $M_2$,  (as indicated in the individual panels). The solid
histograms give the  distributions for a model with efficient common-envelope
(CE) ejection and standard magnetic braking (model 1), while the dashed
histograms show the distributions for inefficient CE ejection and
magnetic braking enhanced by a factor of 5 (model 8). The relative
weights of the different
mass ranges are denoted as $f_{\rm rel}$ in each panel (for model 1 only).
The vertical dashed and dot-dashed curves in the right panels indicate
the range of evolution parameter in which systems may become AM CVn stars.}
\end{figure}
%
% Table 2
%
\begin{table*}
\caption{Galactic birth rates (per year) for CVs, ultra-compact CVs
and CVs with evolved secondaries.}
\tabcolsep=4pt
\begin{tabular}{ccccccccc}
\hline
\hline
\noalign{\vspace{1pt}}
&&&&\multicolumn{3}{c}{$P_{\rm min}$}\\
\noalign{\vspace{2pt}}
\cline{5-7}
\noalign{\vspace{2pt}}
$f_{\rm MB}$&$\alpha_{\rm CE}$&$\alpha_{\rm th}$&total&$< 25$\,min&$<45$\,min
&$<70$\,min& $X_c<0.4$&model\\
\noalign{\vspace{1pt}}
\hline
1&1&1&
6.0E-03& 6.4E-05& 1.2E-04& 6.6E-04& 5.8E-04
&(1)\\
&0.5&0&
9.4E-03& 7.7E-05& 1.8E-04& 1.4E-03& 1.1E-03
&(2)\\
0&1&1&
2.2E-03& 1.1E-04& 1.5E-04& 4.6E-04& 4.8E-04
&(3)\\
&0.5&0&
4.1E-03& 1.9E-04& 2.7E-04& 8.7E-04& 8.8E-04
&(4)\\
0.2&1&1&
4.7E-03& 6.8E-05& 1.2E-04& 6.4E-04& 5.7E-04
&(5)\\
&0.5&0&
8.0E-03& 1.1E-04& 2.1E-04& 1.3E-03& 1.0E-03
&(6)\\
5&1&1&
7.1E-03& 6.4E-05& 1.2E-04& 6.8E-04& 5.6E-04
&(7)\\
&0.5&0&
1.0E-02& 1.4E-05& 6.8E-05& 1.3E-03& 9.4E-04
&(8)\\
1&1&1&
6.1E-03& 7.0E-05& 1.3E-04& 6.8E-04& 6.0E-04
&(1')\\
1&1&1&
7.0E-03& 9.0E-05& 1.5E-04& 9.2E-04& 8.1E-04
&(1$^*$)\\
\noalign{\vspace{2pt}}
\hline
\end{tabular}\\
\parbox{6truein}{
\noindent{Note. ---}
$f_{\rm MB}$: efficiency of magnetic braking;
$\alpha_{\rm CE}$, $\alpha_{\rm th}$: common-envelope efficiency parameters;
$P_{\rm min}$: period minimum; $X_c$: initial central hydrogen abundance
of  the secondary. Models (1) to (8) assume a Galactic age of $12\,$Gyr,
model (1') of 15\,Gyr. Model (1$^*$) is a model similar to model (1)
for a metallicity $Z=0.001$. Note that CVs with He white dwarfs
were not included in these rates.}
\end{table*}

In Figure~2 we illustrate how the CE criterion and the assumptions
concerning the efficiency of magnetic braking affect the distribution
of evolutionary states of the secondary at the beginning of mass transfer.
For this purpose we have defined an evolution parameter $r$ as
\begin{equation}
r={R-R_0\over R_{\rm max} - R_0},
\end{equation}
where $R$ is the radius of the secondary at the beginning of mass transfer,
and  $R_0$ and $R_{\rm max}$ are the radius of the secondary on the
zero-age main sequence and at the point of core hydrogen exhaustion,
respectively; i.e. for a completely unevolved star $r=0$, and for
a star at the end of its main-sequence phase $r=1$.
Figure~2 shows histograms of the evolution parameter for models with
efficient and inefficient CE ejection, respectively, and for different
efficiencies of magnetic braking, $f_{\rm MB} = 0$, 0.2, 1, 5, where
we scaled the efficiency with the standard magnetic-braking description
adopted in our binary evolutionary sequences (see \S~2). Note that in these
distributions we did not include systems with He white dwarf primaries,
since they are not represented in our grid of evolutionary
sequences (although many of these will, of course, also contribute
to the CV population, see e.g. Politano 1996; HNR).

In all cases, the distributions are sharply peaked around $r=0$,
i.e. are dominated by unevolved systems, as required in the standard
explanation for the period gap, and show a tail extending towards long
evolution parameter.  If no magnetic braking is taken into account
($f_{\rm MB}=0$), the distribution with efficient and inefficient CE
ejection (the dashed and dot-dashed histograms at the top) are almost
identical.  As the efficiency of magnetic braking is increased, the tail
with evolved systems decreases, and the distribution becomes even more
concentrated around $r=0$, simply because magnetic braking brings
an increasing number of systems into contact before they had time
to evolve appreciably. The distributions with efficient and inefficient
CE ejection also start to diverge. In the most extreme case, where
magnetic braking was assumed to be enhanced by a factor of 5,
there are no systems with $r>1$ for the model with inefficient CE ejection
(the dot-dashed histogram at the bottom). The reason is that in this model,
the post-CE orbital period distribution has a cut-off for periods above
$\sim 6\,$d, and magnetic  braking
is efficient enough to bring these systems into contact before
any secondary (with $M_2 \le 1.4\Msun$)
had time to evolve off the main sequence.

To further illustrate the dependence of the distributions of evolution
parameter on the mass of the secondary, we divided the distribution
into different mass bins (see Fig.~3), both for our standard
reference model (with $\alpha_{\rm CE}=\alpha_{\rm th}=f_{\rm MB} =1$)
and the extreme model with $\alpha_{\rm CE}= 0.5$, $\alpha_{\rm th}=
0$, $f_{\rm MB} =5$. For masses less than 0.8\Msun, all systems are
relatively unevolved at the beginning of mass transfer, simply because
their masses are too low for them to evolve appreciably in $12\,$Gyr, our
assumed age for the Galaxy. These are the systems that have dominated
in previous population synthesis studies of CVs (see de Kool 1992;
Shafter 1991; Kolb 1993; Politano 1996; HNR). On the other hand, for
more massive secondaries, the distributions contain a substantial
fraction of evolved systems. In the panels on the right, the vertical
dashed line indicates the approximate value for the evolution
parameter (for each mass bin) above which systems will evolve towards
ultra-short periods.  The dot-dashed line gives the approximate
evolution parameter that corresponds to the bifurcation period above
which systems evolve towards long orbital periods rather than short
periods (see e.g. Pylyser \& Savonije 1989 and PRP).  Hence systems
between these two lines are potential candidates for AM CVn
stars. The value of $f_{\rm rel}$ in each panel gives the
fraction of systems in each mass bin (listed only for the standard
case). Note that the number of systems with $M\ge 0.8\Msun$ is roughly
the same as the number of systems with less massive secondaries, which
have dominated in previous CV population synthesis studies.

In Table~2 we present the birthrates in the Galaxy (per year) for the
total population of CVs, for CVs that have minimum periods shorter
than the standard minimum periods (as indicated in the heading),
and CVs with evolved secondaries  (with an initial central hydrogen
abundance $X_c<0.4$) for our eight
main BPS simulations and for two additional simulations where we
assumed an age of $15\,$Gyr and a metallicity of $Z=0.001$
to illustrate the age and metallicity dependence of the results, respectively.
In interpreting this table, two things need to be kept in mind: first,
it does not include CVs with either He white dwarfs or secondaries
with $M< 0.35 \Msun$, since these are not represented in our grid
of evolutionary sequences. Second, since all our evolutionary sequences
assume a magnetic-braking efficiency of 1 (when the secondaries
have a convective envelope and a radiative core) and since the bifurcation
period depends on the magnetic-braking law (see Pylyser \& Savonije
1989 and PRP), only models (1), (2) and (1') can be considered
fully self-consistent.

Table~2 shows that the birthrate for CVs varies quite substantially
for different models from $2\times 10^{-3}\,$yr$^{-1}$ to $10\times
10^{-3}\,$yr$^{-1}$ where the value for our standard model falls
in the middle ($6\times 10^{-3}$\,yr$^{-1}$; model 1). The
qualitative behaviour of
these birthrates is easy to understand, since less efficient CE
ejection produces systems with shorter orbital periods and more
efficient magnetic braking brings them into contact faster. Both effects
increase the birthrates of CVs. These Galactic birthrates can be
converted into local birthrate densities by dividing them by an
effective Galactic volume $5\times 10^{11}\,$pc$^3$; this assumes a
local white-dwarf birthrate of $2\times
10^{-12}\,$pc$^{-3}$\,yr$^{-1}$ (Weidemann 1990), since our
star-formation rate produces roughly 1 white dwarf per year in the
Galaxy.  These numbers therefore `predict' a local birthrate density of
$0.4 - 2 \times 10^{-14}\,$pc$^{-3}$\,yr$^{-1}$ in good agreement
with the observational estimate for the total CV
population by Ritter \& Burkert (1986).

Table~2 also shows that the number of CVs with initially evolved secondaries
and systems that become ultra-compact represent roughly 10 per cent of
the total CV population (see \S~4 for further discussion).

%
% Figure 4
%
\begin{figure*}
\centering
\epsfig{file=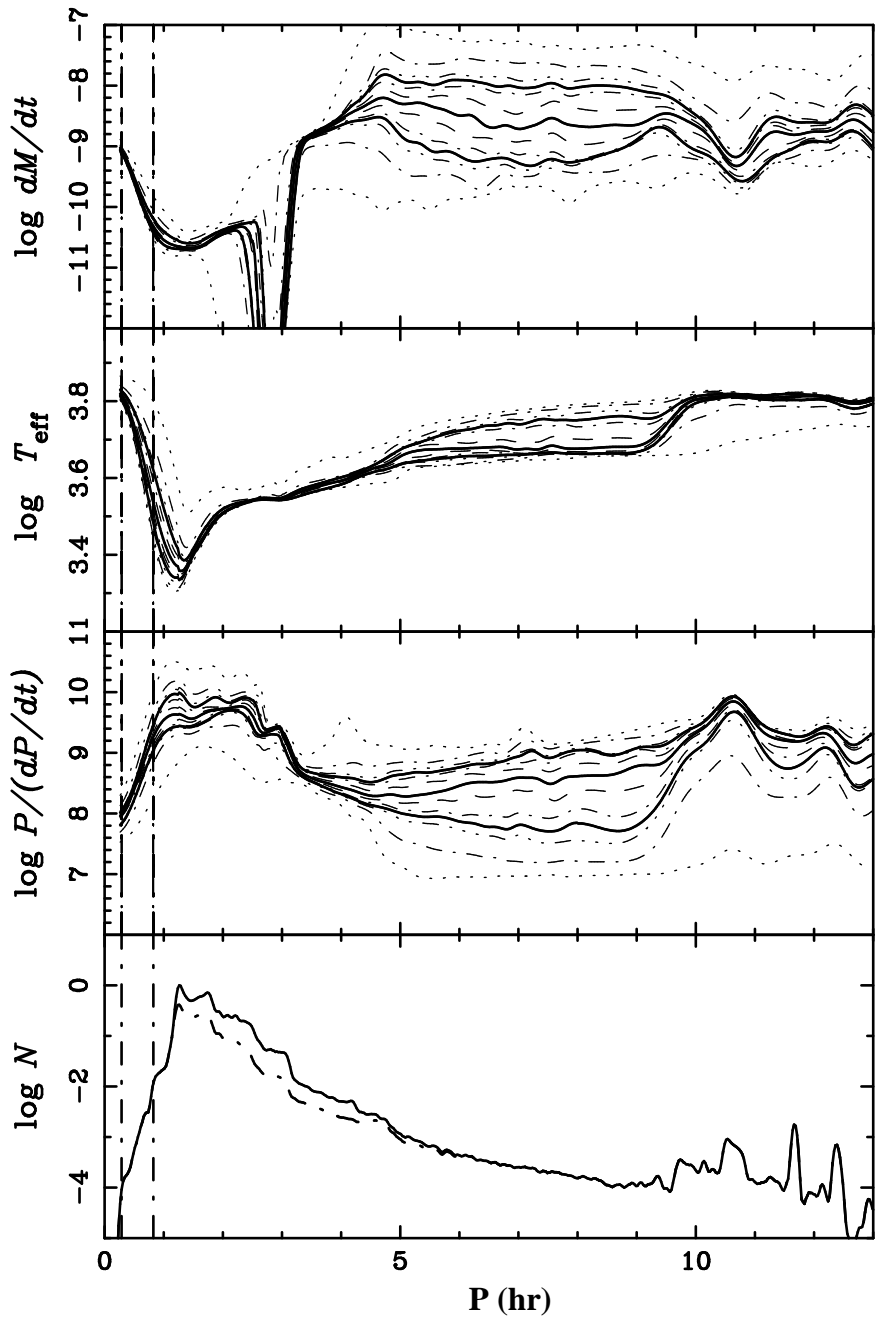,height=\linewidth}
\caption{Intrinsic distributions of mass-transfer rate, $\dot{M}$
(M$_{\odot}$\,yr$^{-1}$), effective temperature, $T_{\rm eff}$ (K),
timescale for orbital period change, $P/|$d$P/$d$t|$ (yr) and total
number distribution as a function of orbital period, $P_{\rm orb}$ for
all systems before they have reached their respective minimum period
(no selection effects applied). In the top three panels, the thick, solid
central curve gives the median of the respective distribution; the pairs
of curves moving progressively outwards from the median include 20, 40, 50,
60, 80 and 98 per cent of the distributions at a given orbital period.
The regions between the thick solid curves contain 50 per cent of all systems
around the median.
In the bottom panel the dot-dashed curve indicates the number of systems
%The shaded regions contain 50 per cent of all systems around the median.
%In the bottom panel the lightly shaded region indicates the number of systems
that initially had evolved secondaries (with $X_c<0.4$) or secondaries
with mass $M_2\ge 1\Msun$. Note that these systems completely dominate
the distribution for orbital periods below
$\sim 1.25\,$hr and above $\sim 4.7\,$hr. All curves have been smoothed
to some degree to reduce the effects of sampling artefacts.}
\end{figure*}

\subsection{Simulations of the properties of the CV population}

To simulate the properties of CVs as a function
of orbital period, we need to know the relative importance of each
of our 120 binary evolution sequences. These sequences span a regular
grid of white dwarf and secondary masses ($M_1$ and $M_2$, respectively) as
well as evolutionary state of the donor, specified by the parameter $r$
(see eq. 2), when the CV mass-transfer phase commences (see \S2 for
details).  We determine these relative weights using the BPS results
discussed above, which gives us the three-dimensional probability
distribution of CVs at the beginning of mass transfer as a function of
$M_1$, $M_2$, and $r$.
For this purpose, we divide this three-dimensional parameter space
into a regular three-dimensional grid and determine the relative probability
for each of these grid elements. For each system to emerge from the BPS
code we locate the nearest 8 grid cells (2 in each dimension) in our
binary evolution sequences and do a tri-linear interpolation to attribute
weights to each of these. For systems with white-dwarf masses
below 0.6\Msun, we use the sequences with $M_1=0.6\Msun$ (note that there are
very few systems below $0.55\Msun$, since
we excluded systems with He white dwarfs). For systems with white-dwarf
masses above 1.0\Msun, we use the sequences with 1\Msun\ (there are also
very few such massive systems). For secondaries with initial masses
below 0.6\Msun, we use the sequences with $M_2=0.6\Msun$, but only consider
the part of the sequence below the appropriate lower mass.

Having determined the relative weights of the 120 binary
evolution sequences, we can then
construct the distributions of the CV properties as a function
of orbital period directly from these sequences, where we also
take into account the time (in yr) a system in each sequence spends
at a particular orbital period.

Figure~4 shows the resulting distributions of the mass-transfer rate
from the secondary $\dot{M}$, its effective temperature, $T_{\rm eff}$,
the timescale for orbital period change, and the total number distribution
as a function of orbital period (note that we included only phases
before the period minimum for clarity). The shaded regions in the top
three panels indicate the range of parameters that include 50 per cent
of all systems at a particular orbital period around the median.
One of the most striking features of these distributions is that
above $\sim 4.7\,$hr they are extremely wide, e.g. spanning several
orders of magnitude in $\dot{M}$. As the shaded area in the bottom panel
shows, at these orbital periods, the distributions are completely
dominated by systems where the secondary was either evolved
initially (with $X_c< 0.4$) or relatively massive (with $M_2\ge 1\Msun$).
The large variation of properties is therefore a direct consequence
of having a mixture of systems with different initial masses and initial
evolutionary states. At a given orbital period, a system with a relatively
more massive secondary will generally have a higher mass-transfer rate,
be more out of thermal equilibrium and evolve more
quickly. On the other hand, for fixed initial masses, the initially more
evolved secondary will have a lower mass-transfer rate at a particular
orbital period (as can be clearly seen in Fig.~1 by comparing $\dot{M}$
at a particular orbital period; see also Beuermann et al.\ [1998]
and Baraffe \& Kolb [2000] for similar discussions). Since this
has the consequence that they evolve more slowly at these periods,
they tend to contribute more to the overall distribution (see \S~4.2).
Thus, for example, although only $\sim 15\%$ of CVs which commence
mass transfer with an orbital period $\ga 5\,$hr, have
evolved secondaries, the longer dwell times (i.e. slower evolution)
makes them by far the most numerous CVs to be found in this period
range.

Note that there is a peak in the overall number distribution just
above 10\,hr. This peak is caused by systems that started mass
transfer near the bifurcation period.  These tend to spend a large
fraction of the evolution in this period range (see the dot-dot-dot-dashed
and dot-dashed curves in Fig.~1) because of the approximate balancing of
opposing effects that drive their period evolution: the angular-momentum
losses due to magnetic braking and gravitational radiation that drive the
systems towards shorter orbital periods and the effects of mass transfer
and nuclear evolution that tend to lead to an increase in the
orbital period (once the mass ratio has been reversed).

Below the orbital period of $4.7\,$hr, all the distributions are
fairly narrow until they reach the standard period minimum ($\sim
75\,$min).  The main reason for this convergence is that below
$4.7\,$hr the distribution is dominated by initially unevolved
systems. As is well known, all evolutionary tracks for normal CVs tend
to converge after the initial turn-on phase or initial phase of
thermal-timescale mass transfer (see e.g. Rappaport et al.\
1983; Stehle, Ritter \& Kolb 1996). Even the systems with initially
evolved secondaries tend to converge towards this universal
track (see Fig.~1), since at these periods the evolution is mainly driven
by angular-momentum loss due to magnetic braking.  Since the
observed distribution of $\dot{M}$ tends to be much wider, this
requires an additional mechanism as, for example, variability induced
by nova events which lead to semi-periodic jumps in the location of
the Roche lobe in the stellar atmosphere of the secondary (see
e.g. Kolb et al.\ 2001, and references therein).  Below the standard
period minimum, the distributions widen again because of a large
variation of the properties of the secondaries at a given period
(e.g. effective temperature, mass, surface abundances).  At the
shortest orbital periods, $\dot{M}$ starts to increase quite sharply
since the timescale for gravitational radiation, which drives the
evolution at this stage, decreases rapidly. Note also that the
temperature distribution is peaked towards higher temperatures just as
the secondaries in individual sequences become hotter (see Fig.~1).

\section{Discussion}

\subsection{The formation of ultra-compact white-dwarf binaries and AM CVn
stars}

A substantial fraction of CVs in our simulations (typically 10 per
cent of the total sample, excluding CVs with He WDs) evolve towards
periods shorter than the classical period minimum. These ultra-compact
CVs are excellent candidates for AM CVn stars. To assess their
relative importance, we need to compare this channel to the
alternative channels discussed in the past, the double-degenerate
channel and the helium-star channel (see \S~1). NPVY have done the
most thorough BPS analysis of these channels to-date, including a
detailed assessment of the uncertainties in the theoretical modelling.
Their estimated Galactic birthrates vary from $0.04 - 4.7\times
10^{-3}\,$yr$^{-1}$ for the double-degenerate model and from $0.9 -
1.6\times 10^{-3}\,$yr$^{-1}$ for the helium-star model (see table~1
in NPVY). In comparison, we obtain birthrates from $0.5 - 1.3 \times
10^{-3} \,$yr$^{-1}$ (see Table~2). These numbers are, however, not
directly comparable. First, while NPVY assume a similar star-formation
rate (by mass) at the present epoch (3.6 versus 3.5\Msun\,yr$^{-1}$),
their star-formation rate was much higher in the past, starting with
a star-formation rate of 15\Msun\,yr$^{-1}$ and decaying with an
exponential decay time of 7\,Gyr (Nelemans et al.\ 2001b). For their
adopted Galactic age of 10\,Gyr, this gives an average star-formation
rate of 8\Msun\,yr$^{-1}$. Since in our simulations, the sample of AM
CVn stars is dominated by relatively old systems, this would suggest
that one should multiply our numbers by a factor of 2 to 4 for the
purpose of comparing the two models (there are several other differences
in the model assumptions which cumulatively make only a minor difference 
of order 1).
On the other hand, only a fraction of our systems (roughly 20 per cent),
reach very short periods ($\la 45\,$min), while all systems in
the study of NPVY are guaranteed to do so.  Considering all the
uncertainties in these various estimates, we think that a fair
conclusion is that the CV channel provides a potentially important
channel for the formation of AM CVn stars, competitive with the
double-degenerate and the helium-star channel, although it may or
may not be the dominant one.

In Table~1 we list a number of model parameters for the six
well-established AM CVn stars, for systems before and after their
period minimum, respectively. The quoted ranges for these parameters
include 90 per cent of all systems around the median at the respective
orbital period, although we note that, at the shortest
orbital periods, these
are almost certainly underestimates since this period range is
not very well sampled in our simulations. In the last two columns we
also list the mass estimates for these 6 systems
from both the double-degenerate (DD) and
the helium-star model (He), again taken directly from NPVY. The
differences in masses can be largely attributed to differences in
composition, the degree of degeneracy and possibly some
differences in the assumed equation of state (implicit
in the approximate mass -- radius relation for degenerate
stars in NPVY). As already noted earlier, one possible way of
distinguishing between the CV channel and the alternative channels for
AM CVn stars is that, in the CV channel, AM CVn stars still have some
hydrogen left in their envelopes when they approach the period
minimum, but very little or none after the period minimum (see column
`$X$' in Table~1). As far as we are aware, of the 6 AM CVn stars in
the table, only one, GP Com, has a sufficiently good spectrum that it
is confirmed to be very hydrogen-deficient; Marsh, Horne \& Rosen
(1991) have determined an upper limit on the ratio of hydrogen to
helium abundance (by number) of H/He$< 10^{-5}$.

Since we are likely to see only a small fraction of the total
population of AM CVn stars, a detailed comparison of the simulations
with observations is seriously compromised by  severe selection
effects (as is the population of ordinary CVs; see e.g. HNR). NPVY
have attempted to model these in a reasonably physical manner. Here,
for the purpose of illustration, we adopt a more {\em ad hoc}
procedure (also see HNR for discussion). To simulate observational
number distributions, we multipy the actual number of systems by an
observing efficiency, $f_{\rm obs}$, which mainly
depends on the mass-transfer
rate $\dot{M}$, i.e., $f_{\rm obs}\propto \dot{M}^\gamma$ (for
$\dot{M}\le 4\times 10^{-10}\Msyr$), where the power in the
exponent increases from 1 (below $\dot{M}\le 4 \times 10^{-10}\Msyr$) to
2 (below $\dot{M}\le 1 \times 10^{-10}\Msyr$), and finally 3 (below
$\dot{M}\le 4 \times 10^{-11}\Msyr$). We also assume that very short
orbital periods are easier to determine and that the
detection efficiency, below an orbital period of 2\,hr, increases
$\propto P_{\rm orb}^{-1}$. We emphasize that this observational
selection criterion is not based on a quantitative physical model, but
was constructed mainly to get (1) a number distribution that has a
similar appearance to the observed one (Ritter \& Kolb 1998), (2)
reduces the spike near the standard period minimum, and (3) shows the
behaviour at ultra-short periods clearly.  The main purpose of this
exercise is to illustrate the dependence of the number of observable
systems on the various model parameters.  Note also that this
selection criterion does not strongly affect systems above the period
gap, which mostly have much higher values for $\dot{M}$ (see Fig.~4)).

%
% Figure 5
%
\begin{figure}
\centering
\epsfig{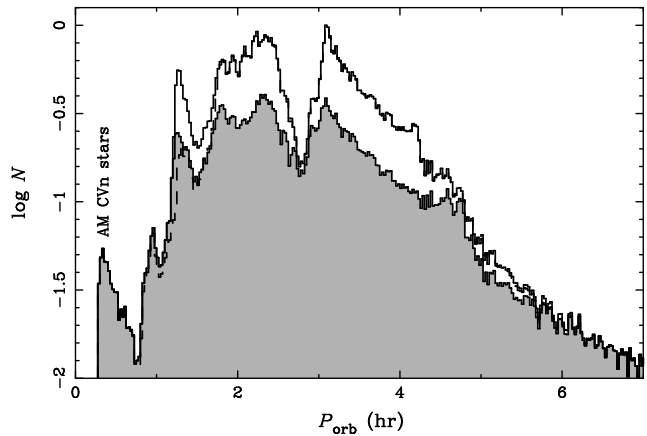}
\caption{Simulated  number distribution for systems as a function of
orbital period with our {\it ad hoc} observational selection criterion for
model (1). The lightly shaded area shows the distribution of systems
that started either with an evolved secondary (with $X_c <0.4$) or a
secondary of mass $\ge 1$\.\Msun.  The dashed histogram, only distinct
around an orbital period of 1.25\,hr, shows the distribution where
systems with $\dot{M}\le 2.5\times 10^{-11}\Msun\,$yr$^{-1}$ were
eliminated from the observable sample.  Note that systems with $P_{\rm
orb} \ga 4.7$ hr and $P_{\rm orb} \la 1.25$ hr are largely ones that
started with evolved or more massive secondaries.  The region of the
simulated AM CVn systems is indicated on the figure.}
\end{figure}

%
% Table 3
%
\begin{table*}
\caption{Predicted number distributions for CVs with CO white dwarfs
with our {\em ad hoc} selection criterion (normalized to the total of
$\sim400$ currently known systems).}

\tabcolsep=4pt
\begin{tabular}{ccccccccccc}
\hline
\hline
\noalign{\vspace{1pt}}
&&&\multicolumn{7}{c}{$P$\,(hr)}\\
\noalign{\vspace{2pt}}
\cline{4-10}
\noalign{\vspace{2pt}}
$f_{\rm MB}$&$\alpha_{\rm CE}$&$\alpha_{\rm th}$&0\,--\,0.42&0.42\,--\,0.75
&0.75\,--\,1.17&1.17\,--\,2.5&2.5\,--\,3.0&3.0\,--\,5.0&$>5.0$& model\\

\noalign{\vspace{1pt}}
\hline
1&1&1&
1.7 &   1.9 &   5.1 & 173 &  37 & 156 &  25
& (1)\\
&0.5&0&
1.1 &   1.5 &   6.5 & 176 &  36 & 160 &  19
& (2)\\
0&1&1&
7.6 &   6.3 &  10.4 & 162 &  40 & 139 &  34
& (3)\\
&0.5&0&
8.5 &   6.8 &  11.3 & 163 &  37 & 142 &  33
& (4)\\
0.2&1&1&
2.2 &   2.4 &   6.2 & 170 &  39 & 153 &  27
& (5)\\
&0.5&0&
2.3 &   2.3 &   7.5 & 174 &  37 & 156 &  21
& (6)\\
5&1&1&
1.4 &   1.5 &   4.2 & 175 &  35 & 159 &  24
& (7)\\
&0.5&0&
0.0 &   0.6 &   4.6 & 180 &  36 & 164 &  15
& (8)\\
1&1&1&
1.7 &   1.8 &   5.1 & 173 &  37 & 156 &  25
& (1')\\
1&1&1&
1.4 &   1.8 &   4.7 & 165 &  44 & 149 &  34
& (1$^*$)\\
\noalign{\vspace{2pt}}
\hline
\end{tabular}\\
\end{table*}

Figure~5 shows the simulated number distribution with this selection
criterion for model (1) where the lightly shaded region indicates the
distribution of systems that started either with an initially evolved
secondary (with $X_c < 0.4$) or a secondary of mass $\ge 1\Msun$.  As
already discussed in \S~3.3, these evolved/massive systems completely
dominate the distribution below $\sim 1.25\,$hr and above $\sim
4.7\,$hr. While they also fill in part of the period gap (between
$\sim 2.5$ and $\sim 3\,$hr in our evolution model; see \S~2), the
period gap still remains well defined. (Note, however, that CVs with
He white dwarfs and CVs that start mass transfer in the orbital period
range will further increase the number of systems in the period gap;
see e.g. HNR). Around an orbital period
of 75\,min, the distribution shows the well-known period-minimum spike
(since the orbital period evolves slowly near the period minimum).
This spike is not apparent in the observed distribution, perhaps suggesting
more drastic selection effects (see the detailed discussion of
this problem in Kolb \& Baraffe 1999).  To illustrate this
possibility, we performed another simulation where we assumed that no
systems with $\dot{M}$ below $2.5\times 10^{-10}\Msyr$ were detectable.
This simulation is shown as a dashed histogram in Figure~5 and is only
clearly visible near the period minimum. The only significant change
this additional assumption introduces is that it more or less
completely eliminates the period-minimum spike. However, this requires
significant fine-tuning because the range of mass-transfer rates near
the period minimum is very small, and therefore this cannot be considered
very plausible. On the other hand, if there were a larger range of $\dot{M}$
near the period minimum or a larger variation of the minimum period
(which is not present in current evolutionary models), one can imagine
that the spike would be spread out. A more drastic alternative,
suggested by King (2001), would be that CVs never reach the
minimum period in the lifetime of the Galaxy, in which case the
observed cut-off near 80\,min is an age effect.  However, this suggestion
is not consistent with standard BPS assumptions as employed in the
present and other related studies and would require some drastic alterations
to the entire BPS model.

Below the minimum period, there is a further spike in the orbital-period
distribution around 20\,min, the period range where most AM CVn stars
are observed. Even though the number of actual systems decreases
monotonically with decreasing orbital period, the mass-transfer rates
go up sharply, which makes the systems with ultra-short periods
easier to detect observationally. This `AM CVn' spike is therefore
very sensitive to the observational selection criterion (it is also
present in the study of NPVY who use a more physical model for
the observational detectibility of their systems).

In Table~3 we present the number distribution in various period ranges
for all of the BPS models using our {\em ad hoc} selection criterion,
normalized to a total number of 400 systems (approximately the number
of known CVs). This illustrates the variation in the number of
observable systems with the BPS parameters.  It shows that the number
of observable AM CVn stars (resulting only from the CV channel) is
largest if there is no magnetic braking before the onset of mass
transfer and decreases as the efficiency of magnetic braking is
increased (see the discussion in \S~3.2). There is generally a good
correlation between the number of systems with periods above 5\,hr
(dominated by evolved/massive systems), systems in the period gap, and
ultra-compact systems.

Finally we note that the number distributions are also somewhat sensitive to
other factors which we did not vary systematically, in particular the
assumed age of the Galaxy and the metallicity of the secondaries.  Since AM
CVn stars in our model originate from systems where the secondaries
start mass transfer near the end of, or just after, their main-sequence
phase, the age of the Galaxy determines the lowest initial mass of a
secondary that can evolve to the AM CVn stage. Similarly, since
low-metallicity stars evolve much more rapidly, population-II stars of
lower mass will be able to evolve to become ultra-compact systems. In
Tables~2 and 3, the models (1') and (1$^*$) give the birthrates
and number distributions for models with a Galactic age of
15\,Gyr and for a population II metallicity with $Z=0.001$ (for an age
of 12\,Gyr), where all other parameters were the same as in model (1).
simulations.
%Note, in particular, how
%the birthrates for CVs and AM CVn stars increase sharply in the
%low-metallicity simulation. Interestingly, Marsh et al.\ (1991) have
%made the suggestion based on their spectroscopic analysis that GP Com is a
%low-metallicity halo object.

%
% Figure 6
%
\begin{figure*}
\centering
\epsfig{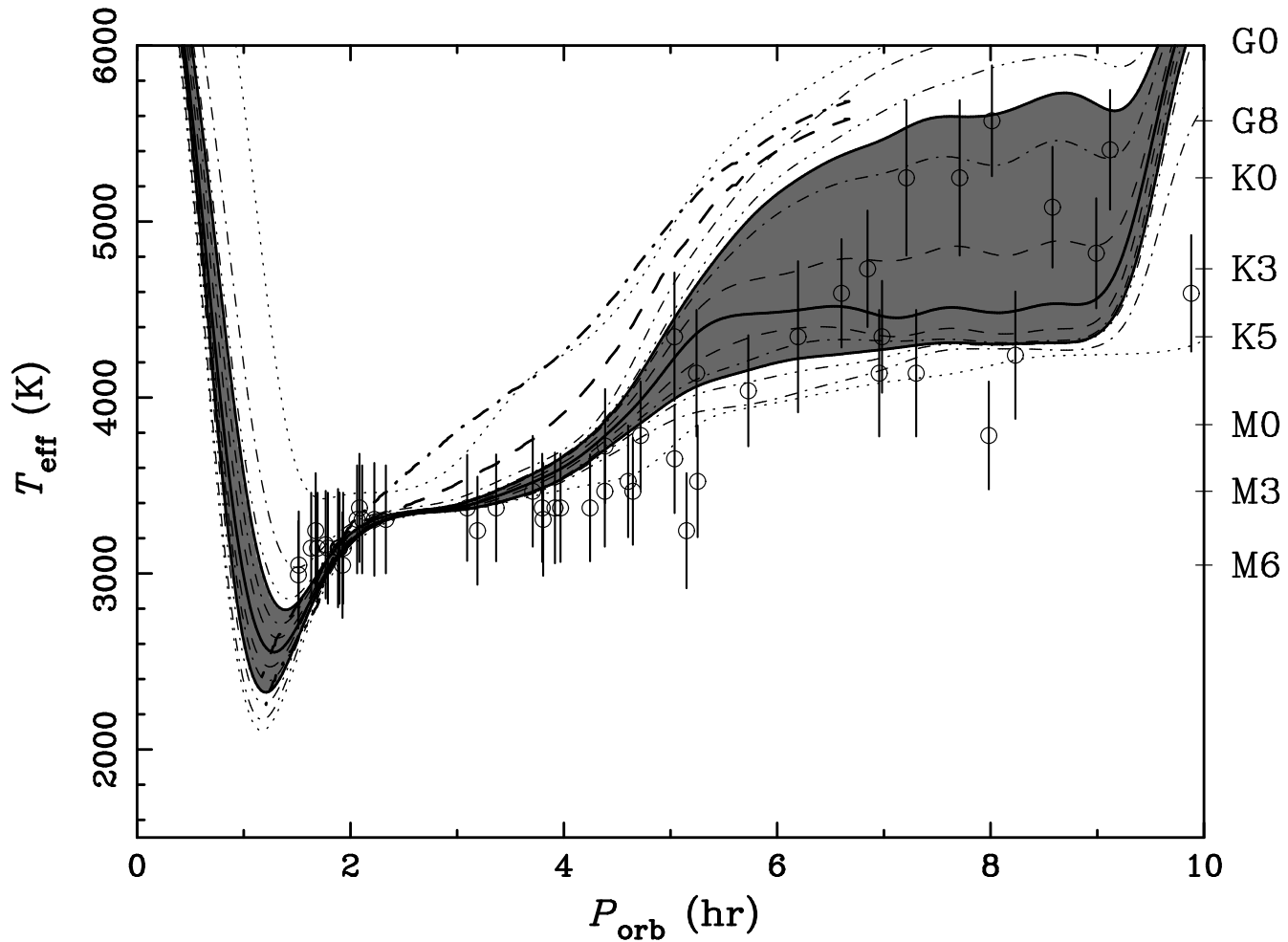}
\caption{Comparison of the theoretical distribution of effective
temperatures/spectral type as a function of orbital period with the
observational data of Beuermann et al.\ (1998). The various curves
show the intrinsic distribution of temperatures for model (1) 
(as in Fig.~4),
where the shaded region includes 50 per cent of all systems around the
median (only for systems before they reached the period minimum). The
circles with error bars give the temperatures of systems with spectral
types taken from Beuermann et al.\ (1998).  The thick dashed curve
shows the temperature of systems that are unevolved and in thermal
equilibrium if they fill their Roche lobes at a particular orbital
period (assuming a white dwarf of 0.7\,Msun and with temperature
corrections applied). The dot-dashed curve shows the temperature of
the same systems without temperature corrections applied (see text for
details).}
\end{figure*}

\subsection{Cataclysmic variables with evolved companions}

In Figure~6 we present a comparison of our simulated distribution of the
effective temperature of the secondary as a function of orbital period
with the spectral types of secondaries
from the sample of Beuermann et al.\ (1998). To plot the observed
systems, we used the spectral type -- effective temperature relation
of Schmidt-Kaler (1982) and added (in quadrature) an error of 300\,K,
due to the uncertainties in this relation at late spectral types, to
the errors in the spectral-type determination given by Beuermann et
al.\ (1998). In addition, since our stellar models tend to
overestimate the effective temperature below $\sim 4500\,$K compared
to evolutionary calculations with a more sophisticated treatment of
the stellar atmosphere (see Chabrier \& Baraffe 1997; Baraffe et al.\
1998), we applied corrections to our effective temperatures to make
the comparison more realistic. These corrections were obtained by
comparing the effective temperatures of unevolved stars between 0.1
and 1\Msun\ in our models with the models of Baraffe et al.\ (1998). The
difference between the dashed and the dot-dashed curves in Figure~6
shows the magnitude of this correction for unevolved stars in thermal
equilibrium, which start to fill their Roche lobes at a particular orbital
period (the correction has a maximum of 350\,K around
4000\,K)\footnote{We note that the determination of this correction is
not entirely straightforward, since the Baraffe et al.\ (1998) models
have not been calibrated to reproduce a solar model in a
self-consistent way.}.

As Figure~6 shows, our simulations reproduce the observed distribution
reasonably well at all orbital periods. In particular, they show a
large variation of effective temperatures/spectral types at orbital
periods longer than $\sim 5\,$hr, consistent with the observed
variation and confirming the suggestion by Beuermann et al.\ (1998)
and Baraffe \& Kolb (2000) that systems at these periods are dominated
by those with evolved secondaries. Many of these systems may therefore
show evidence for nuclear processing (in particular CNO processing)
in their envelopes. Below an orbital period of $\sim
2\,$hr, the observed secondaries are close to the relation for
unevolved stars in thermal equilibrium (the thick dashed curve) as
expected in the standard model for CVs, where the secondaries remain
very close to thermal equilibrium until near the end of their
hydrogen-burning phase.
Near an orbital period of $\sim 5\,$hr, the observed spectral types
appear to be slightly, but systematically later than our simulations
imply. The reason may be that the applied effective temperature
corrections were not large enough or that our stars are not
sufficiently out of thermal equilibrium in this period range (perhaps
consistent with having a period gap that is slightly too small; see
\S~2; for a different interpretation see Baraffe \& Kolb 2000). We
also note that for orbital periods longer than $\sim 5\,$hr, some of
the observed systems could originate from more massive, supersoft
systems (not included in our simulations), which under certain
conditions may reach orbital periods as short as $4\,$hr (King et al.\
2001) before the period increases again. However, since this occurs
during a fast evolutionary phase with large $\dot{M}$, they are
unlikely to dominate the distribution below orbital periods of $\sim
10\,$hr, which in our simulations are dominated by systems with
relatively low $\dot{M}$ with evolved secondaries (see Figures~1 and
4).

\section{Conclusion}

We have presented a self-consistent study of CVs and CV-like systems,
combining detailed binary population synthesis simulations with
a large grid of binary evolutionary calculations, where we included
systems with evolved secondaries up to a mass of 1.4\Msun.
These simulations show that CVs above $\sim 5\,$hr are dominated
by evolved, relatively massive systems and that the predicted range
in effective temperatures is consistent with the variation in
spectral types obtained by Beuermann et al.\ (1998). These systems,
which constitute $\sim 10$ per cent of the total CV population,
are also the progenitors of ultra-compact systems with periods
less than $\sim 1\,$hr which are good candidates for AM CVn stars. 
For standard population synthesis assumptions, we obtain AM CVn
birthrates that are competitive with those in alternative
models. While these systems do not experience a period gap between 2
and 3\,hr, their contribution is sufficiently small that they do not
destroy its distinct appearance.

\section*{Acknowledgements}

This work was in part supported by a Royal Society UK-China Joint Project
Grant (Ph.P and Z.H.), the Chinese National Science Foundation under 
Grant No.\ 19925312, 10073009 and NKBRSF No. 19990754 (Z.H.) and by the 
National Aeronautics and Space Administration under ATP grant NAG5-8368.

\appendix

\section{Appendix}

Table~A1 lists some of the parameters which are important in the present
study for the 120 sequences in our grid. These include the initial
parameters, parameters at the period minimum and the size of the
period gap, if present. (For the definitions of the parameters,
see the bottom of the table.) The last column gives the relative
weight of each sequence in the CV population synthesis for BPS model
(1).

%
% Table A1
%
\begin{table*}
\caption{Selected Properties of Binary Sequences}
\tabcolsep=5pt
\begin{tabular}{ccrcccccccccc}
\hline
\hline
\noalign{\vspace{1pt}}
\multicolumn{6}{c}{Initial Parameters}&&\multicolumn{4}
{c}{Parameters at Period Minimum}\\
\noalign{\vspace{2pt}}
\cline{1-6}\cline{8-11}
\noalign{\vspace{2pt}}
$M_1$&$M_2$&\multicolumn{1}{c}{$P^i$}&$t^i$&$X_c^i$&$M_c^i/M$&&
$P_{\rm min}$&$M_2^{\rm min}$&$\dot{M}_{\rm min}$& $X_s^{\rm min}$&
gap&weight\\
(\Msun)&(\Msun)&\multicolumn{1}{c}{(hr)}&(yr)&&&&(min)&(\Msun)&(\Msyr)&&(hr)\\
\noalign{\vspace{2pt}}
\hline
\noalign{\vspace{2pt}}
 0.6 &  0.6 &   4.1 &  0.0E+00 &  0.685 &  0.000 &  &
 74.8 &  0.069 &  2.1E$-$11 &  0.682 &
 2.52$-$3.08 &    1.0E$-$01\\
\noalign{\vspace{5pt}}
 0.6 &  0.8 &   5.3 &  0.0E+00 &  0.685 &  0.000 &  &
 75.2 &  0.067 &  2.0E$-$11 &  0.682 &
 2.46$-$3.07 &    5.0E$-$02\\
 0.6 &  0.8 &   5.5 &  3.9E+09 &  0.560 &  0.000 &  &
 74.1 &  0.063 &  1.9E$-$11 &  0.662 &
 2.33$-$2.92 &    4.6E$-$02\\
 0.6 &  0.8 &   5.7 &  8.0E+09 &  0.436 &  0.000 &  &
 73.3 &  0.056 &  1.6E$-$11 &  0.638 &
 1.75$-$2.07 &    9.5E$-$03\\
\noalign{\vspace{5pt}}
 0.6 &  0.9 &   5.8 &  0.0E+00 &  0.685 &  0.000 &  &
 75.2 &  0.066 &  2.1E$-$11 &  0.682 &
 2.43$-$3.07 &   3.6E$-$02\\
 0.6 &  0.9 &   6.1 &  2.7E+09 &  0.552 &  0.000 &  &
 73.8 &  0.066 &  2.1E$-$11 &  0.659 &
 2.31$-$2.87 &    2.5E$-$02\\
 0.6 &  0.9 &   6.4 &  5.1E+09 &  0.433 &  0.000 &  &
 72.9 &  0.068 &  2.1E$-$11 &  0.635 &
 1.89$-$2.25 &   6.7E$-$03\\
 0.6 &  0.9 &   6.9 &  7.7E+09 &  0.310 &  0.000 &  &
 70.6 &  0.055 &  1.9E$-$11 &  0.594 &
 1.35$-$1.72 &    5.5E$-$03\\
 0.6 &  0.9 &   8.1 &  1.2E+10 &  0.093 &  0.000 &  &
 64.7 &  0.045 &  1.6E$-$11 &  0.491 &
 1.22$-$1.45 &   2.1E$-$03\\
\noalign{\vspace{5pt}}
 0.6 &  1.0 &   6.5 &  0.0E+00 &  0.685 &  0.000 &  &
 75.3 &  0.067 &  2.1E$-$11 &  0.682 &
 2.45$-$3.07 &   3.0E$-$02\\
 0.6 &  1.0 &   6.9 &  1.5E+09 &  0.567 &  0.000 &  &
 73.8 &  0.066 &  2.1E$-$11 &  0.660 &
 2.24$-$2.90 &    2.6E$-$02\\
 0.6 &  1.0 &   7.3 &  3.2E+09 &  0.444 &  0.000 &  &
 72.8 &  0.060 &  1.9E$-$11 &  0.635 &
 1.93$-$2.35 &    7.3E$-$03\\
 0.6 &  1.0 &   7.8 &  4.6E+09 &  0.327 &  0.000 &  &
 62.7 &  0.046 &  1.3E$-$11 &  0.574 &
 1.31$-$1.44 &   6.1E$-$03\\
 0.6 &  1.0 &   8.9 &  7.1E+09 &  0.101 &  0.000 &  &
 54.8 &  0.041 &  1.5E$-$11 &  0.405 &
 1.30$-$1.45 &    5.2E$-$03\\
 0.6 &  1.0 &  10.8 &  9.0E+09 &  0.000 &  0.000 &  &
 44.6 &  0.031 &  1.4E$-$11 &  0.196 &
  &     3.4E$-$03\\
 0.6 &  1.0 &  13.2 &  1.0E+10 &  0.000 &  0.037 &  &
 22.1 &  0.065 &  5.8E$-$10 &  0.053 &
  &    1.8E$-$03\\
 0.6 &  1.0 &  15.4 &  1.1E+10 &  0.000 &  0.063 &  &
 11.1 &  0.100 &  4.7E$-$09 &  0.037 &
  &     9.9E$-$04\\
 0.6 &  1.0 &  18.2 &  1.1E+10 &  0.000 &  0.093 &  & \\
\noalign{\vspace{5pt}}
 0.6 &  1.1 &   7.4 &  0.0E+00 &  0.685 &  0.000 &  &
 75.2 &  0.067 &  2.0E$-$11 &  0.682 &
 2.44$-$3.08 &    2.8E$-$02\\
 0.6 &  1.1 &   7.9 &  1.1E+09 &  0.558 &  0.000 &  &
 74.0 &  0.062 &  1.9E$-$11 &  0.657 &
 2.32$-$2.90 &    2.7E$-$02\\
 0.6 &  1.1 &   8.3 &  2.1E+09 &  0.448 &  0.000 &  &
 73.4 &  0.063 &  2.2E$-$11 &  0.634 &
 1.99$-$2.40 &    6.4E$-$03\\
 0.6 &  1.1 &   8.8 &  2.9E+09 &  0.340 &  0.000 &  &
 71.0 &  0.060 &  2.0E$-$11 &  0.604 &
 1.41$-$1.79 &    5.9E$-$03\\
 0.6 &  1.1 &   9.8 &  4.3E+09 &  0.105 &  0.000 &  &
 67.3 &  0.048 &  1.5E$-$11 &  0.537 &
 1.38$-$1.67 &   5.7E$-$03\\
 0.6 &  1.1 &  11.6 &  5.6E+09 &  0.000 &  0.000 &  &
 53.2 &  0.033 &  1.2E$-$11 &  0.265 &
  &     5.1E$-$03\\
 0.6 &  1.1 &  15.4 &  7.1E+09 &  0.000 &  0.039 &  &
 17.8 &  0.076 &  1.4E$-$09 &  0.030 &
  &     3.3E$-$03\\
 0.6 &  1.1 &  18.2 &  7.5E+09 &  0.000 &  0.070 &  & \\
\noalign{\vspace{5pt}}
 0.6 &  1.2 &   8.4 &  0.0E+00 &  0.685 &  0.000 &  &
 75.8 &  0.071 &  2.4E$-$11 &  0.682 &
 2.45$-$3.07 &    2.1E$-$02\\
 0.6 &  1.2 &   9.4 &  1.4E+09 &  0.567 &  0.000 &  &
 71.4 &  0.071 &  2.1E$-$11 &  0.632 &
 1.83$-$2.26 &    5.7E$-$02\\
 0.6 &  1.2 &  10.6 &  2.7E+09 &  0.448 &  0.000 &  &
 66.2 &  0.046 &  1.6E-11 &  0.518 &
 1.73$-$2.05 &    1.7E$-$02\\
 0.6 &  1.2 &  12.0 &  3.7E+09 &  0.329 &  0.000 &  &
 67.3 &  0.047 &  1.5E$-$11 &  0.523 &
 1.79$-$2.05 &    1.9E$-$02\\
 0.6 &  1.2 &  14.9 &  5.2E+09 &  0.105 &  0.000 &  &
 50.5 &  0.034 &  1.6E$-$11 &  0.218 &
 1.33$-$1.35 &    5.9E$-$03\\
 0.6 &  1.2 &  15.6 &  5.4E+09 &  0.047 &  0.000 &  &
 34.7 &  0.042 &  6.4E$-$11 &  0.078 &
  &     1.6E$-$03\\
 0.6 &  1.2 &  16.4 &  5.6E+09 &  0.010 &  0.000 &  & \\
\noalign{\vspace{5pt}}
 0.6 &  1.4 &  10.1 &  1.6E+07 &  0.681 &  0.000 &  &
 75.7 &  0.063 &  1.9E$-$11 &  0.679 &
 2.41$-$3.07 &    9.2E$-$03\\
 0.6 &  1.4 &  11.7 &  9.5E+08 &  0.559 &  0.000 &  &
 72.3 &  0.070 &  2.4E$-$11 &  0.613 &
 2.32$-$2.86 &   3.3E$-$02\\
 0.6 &  1.4 &  13.4 &  1.7E+09 &  0.448 &  0.000 &  &
 68.9 &  0.048 &  1.4E$-$11 &  0.544 &
 2.20$-$2.63 &    1.1E$-$02\\
 0.6 &  1.4 &  15.5 &  2.2E+09 &  0.340 &  0.000 &  &
 55.8 &  0.043 &  2.0E$-$11 &  0.378 &
 2.01$-$2.16 &    6.2E$-$03\\
 0.6 &  1.4 &  18.1 &  2.7E+09 &  0.221 &  0.000 &  &
 47.4 &  0.035 &  2.2E$-$11 &  0.249 &
 1.16$-$1.25 &   4.0E$-$03\\
 0.6 &  1.4 &  20.3 &  3.1E+09 &  0.112 &  0.000 &  &
 40.5 &  0.046 &  3.5E$-$11 &  0.109 &
  &     2.6E$-$03\\
 0.6 &  1.4 &  22.3 &  3.3E+09 &  0.024 &  0.000 &  & \\
\noalign{\vspace{2pt}}
\hline
\end{tabular}\\
\end{table*}
\vfill\eject
\begin{table*}
\contcaption{}
\tabcolsep=5pt
\begin{tabular}{ccrcccccccccc}
\hline
\hline
\noalign{\vspace{1pt}}
\multicolumn{6}{c}{Initial Parameters}&&\multicolumn{4}
{c}{Parameters at Period Minimum}\\
\noalign{\vspace{2pt}}
\cline{1-6}\cline{8-11}
\noalign{\vspace{2pt}}
$M_1$&$M_2$&\multicolumn{1}{c}{$P^i$}&$t^i$&$X_c^i$&$M_c^i/M$&&
$P_{\rm min}$&$M_2^{\rm min}$&$\dot{M}_{\rm min}$& $X_s^{\rm min}$&
gap&weight\\
(\Msun)&(\Msun)&\multicolumn{1}{c}{(hr)}&(yr)&&&&(min)&(\Msun)&(\Msyr)&&(hr)\\
\noalign{\vspace{2pt}}
\hline
\noalign{\vspace{2pt}}
 0.8 &  0.6 &   4.2 &  0.0E+00 &  0.685 &  0.000 &  &
 77.1 &  0.068 &  2.6E$-$11 &  0.683 &
 2.64$-$3.03 &  2.0E$-$02\\
\noalign{\vspace{5pt}}
 0.8 &  0.8 &   5.4 &  0.0E+00 &  0.685 &  0.000 &  &
 75.8 &  0.068 &  2.4E$-$11 &  0.682 &
 2.58$-$3.02 &   1.0E$-$02\\
 0.8 &  0.8 &   5.6 &  3.9E+09 &  0.560 &  0.000 &  &
 75.0 &  0.068 &  2.4E$-$11 &  0.663 &
 2.45$-$2.89 &   8.9E$-$03\\
 0.8 &  0.8 &   5.9 &  8.0E+09 &  0.436 &  0.000 &  &
 75.3 &  0.064 &  2.4E$-$11 &  0.642 &
 2.15$-$2.50 &    1.5E$-$03\\
\noalign{\vspace{5pt}}
 0.8 &  0.9 &   6.0 &  0.0E+00 &  0.685 &  0.000 &  &
 76.0 &  0.068 &  2.4E$-$11 &  0.682 &
 2.59$-$3.02 &    7.9E$-$03\\
 0.8 &  0.9 &   6.3 &  2.7E+09 &  0.552 &  0.000 &  &
 75.3 &  0.064 &  2.2E$-$11 &  0.660 &
 2.44$-$2.88 &    4.6E$-$03\\
 0.8 &  0.9 &   6.7 &  5.1E+09 &  0.433 &  0.000 &  &
 75.0 &  0.064 &  2.5E$-$11 &  0.638 &
 2.08$-$2.45 &    9.8E$-$04\\
 0.8 &  0.9 &   7.1 &  7.7E+09 &  0.310 &  0.000 &  &
 72.4 &  0.064 &  2.3E$-$11 &  0.610 &
 1.67$-$1.87 &    7.6E$-$04\\
 0.8 &  0.9 &   8.4 &  1.2E+10 &  0.093 &  0.000 &  &
 61.1 &  0.050 &  2.4E$-$11 &  0.399 &
 1.19$-$1.43 &    3.6E$-$04\\
\noalign{\vspace{5pt}}
 0.8 &  1.0 &   6.8 &  0.0E+00 &  0.685 &  0.000 &  &
 76.6 &  0.069 &  2.8E$-$11 &  0.682 &
 2.55$-$3.02 &    7.0E$-$03\\
 0.8 &  1.0 &   7.1 &  1.5E+09 &  0.567 &  0.000 &  &
 75.3 &  0.068 &  2.7E$-$11 &  0.661 &
 2.43$-$2.88 &   4.4E$-$03\\
 0.8 &  1.0 &   7.5 &  3.2E+09 &  0.444 &  0.000 &  &
 73.8 &  0.063 &  2.3E$-$11 &  0.637 &
 2.08$-$2.44 &    1.1E$-$03\\
 0.8 &  1.0 &   8.0 &  4.6E+09 &  0.327 &  0.000 &  &
 69.5 &  0.075 &  2.8E$-$11 &  0.590 &
 1.35$-$1.56 &    9.2E$-$04\\
 0.8 &  1.0 &   9.3 &  7.1E+09 &  0.101 &  0.000 &  &
 66.1 &  0.042 &  1.5E$-$11 &  0.461 &
 1.48$-$1.59 &    8.9E$-$04\\
 0.8 &  1.0 &  11.2 &  9.0E+09 &  0.000 &  0.000 &  &
 53.8 &  0.029 &  1.3E$-$11 &  0.255 &
  &    4.0E$-$04\\
 0.8 &  1.0 &  13.6 &  1.0E+10 &  0.000 &  0.036 &  &
 23.2 &  0.068 &  7.2E$-$10 &  0.104 &
  &    2.1E$-$04\\
 0.8 &  1.0 &  15.9 &  1.1E+10 &  0.000 &  0.063 &  &
 10.4 &  0.110 &  1.4E$-$08 &  0.010 &
  &    1.5E$-$04\\
 0.8 &  1.0 &  19.0 &  1.1E+10 &  0.000 &  0.093 &  & \\
\noalign{\vspace{5pt}}
 0.8 &  1.1 &   7.7 &  0.0E+00 &  0.685 &  0.000 &  &
 76.2 &  0.067 &  2.5E$-$11 &  0.682 &
 2.52$-$3.01 &    7.5E$-$03\\
 0.8 &  1.1 &   8.2 &  1.1E+09 &  0.558 &  0.000 &  &
 75.0 &  0.066 &  2.4E$-$11 &  0.658 &
 2.40$-$2.86 &   4.0E$-$03\\
 0.8 &  1.1 &   8.6 &  2.1E+09 &  0.448 &  0.000 &  &
 74.8 &  0.065 &  2.5E$-$11 &  0.637 &
 2.19$-$2.52 &    8.9E$-$04\\
 0.8 &  1.1 &   9.1 &  2.9E+09 &  0.340 &  0.000 &  &
 72.9 &  0.062 &  2.3E$-$11 &  0.612 &
 1.75$-$2.01 &    8.1E$-$04\\
 0.8 &  1.1 &  10.2 &  4.3E+09 &  0.105 &  0.000 &  &
 65.3 &  0.055 &  2.5E$-$11 &  0.457 &
 1.25$-$1.54 &    8.6E$-$04\\
 0.8 &  1.1 &  12.0 &  5.6E+09 &  0.000 &  0.000 &  &
 59.7 &  0.034 &  1.2E$-$11 &  0.333 &
  &     6.2E$-$04\\
 0.8 &  1.1 &  16.0 &  7.1E+09 &  0.000 &  0.038 &  &
 16.7 &  0.079 &  2.0E$-$09 &  0.026 &
  &     4.2E$-$04\\
 0.8 &  1.1 &  18.9 &  7.5E+09 &  0.000 &  0.070 &  & \\
\noalign{\vspace{5pt}}
 0.8 &  1.2 &   8.8 &  0.0E+00 &  0.685 &  0.000 &  &
 76.4 &  0.071 &  2.9E$-$11 &  0.682 &
 2.53$-$3.01 &    9.7E$-$03\\
 0.8 &  1.2 &   9.7 &  1.4E+09 &  0.567 &  0.000 &  &
 74.3 &  0.065 &  2.6E$-$11 &  0.635 &
 2.01$-$2.38 &    7.5E$-$03\\
 0.8 &  1.2 &  11.0 &  2.7E+09 &  0.447 &  0.000 &  &
 70.1 &  0.052 &  1.8E-11 &  0.551 &
 1.32$-$1.40 &    1.9E$-$03\\
 0.8 &  1.2 &  12.5 &  3.7E+09 &  0.329 &  0.000 &  &
 63.2 &  0.042 &  1.6E$-$11 &  0.416 &
 1.88$-$2.23 &    2.4E$-$03\\
 0.8 &  1.2 &  15.5 &  5.2E+09 &  0.105 &  0.000 &  &
 46.7 &  0.049 &  4.8E$-$11 &  0.175 &
  &     7.3E$-$04\\
 0.8 &  1.2 &  16.3 &  5.4E+09 &  0.047 &  0.000 &  &
 28.6 &  0.067 &  4.0E$-$10 &  0.041 &
  &     2.1E$-$04\\
 0.8 &  1.2 &  17.0 &  5.6E+09 &  0.010 &  0.000 &  & \\
\noalign{\vspace{5pt}}
 0.8 &  1.4 &  10.5 &  2.1E+07 &  0.680 &  0.000 &  &
 76.9 &  0.068 &  2.6E$-$11 &  0.673 &
 2.51$-$3.00 &    5.7E$-$03\\
 0.8 &  1.4 &  12.1 &  1.0E+09 &  0.549 &  0.000 &  &
 72.8 &  0.064 &  2.6E$-$11 &  0.589 &
 1.81$-$2.11 &    4.2E$-$03\\
 0.8 &  1.4 &  14.0 &  1.7E+09 &  0.437 &  0.000 &  &
 69.2 &  0.060 &  2.6E$-$11 &  0.522 &
 2.19$-$2.49 &    1.2E$-$03\\
 0.8 &  1.4 &  16.1 &  2.2E+09 &  0.340 &  0.000 &  &
 66.5 &  0.049 &  2.0E$-$11 &  0.471 &
 2.06$-$2.29 &    8.7E$-$04\\
 0.8 &  1.4 &  18.8 &  2.7E+09 &  0.221 &  0.000 &  &
 60.4 &  0.049 &  2.5E$-$11 &  0.358 &
 1.79$-$1.87 &    5.5E$-$04\\
 0.8 &  1.4 &  21.3 &  3.1E+09 &  0.112 &  0.000 &  &
 40.0 &  0.050 &  8.0E$-$11 &  0.104 &
  &     3.2E$-$04\\
 0.8 &  1.4 &  23.3 &  3.3E+09 &  0.023 &  0.000 &  & \\
\noalign{\vspace{2pt}}
\hline
\end{tabular}\\
\end{table*}
\vfill\eject
\begin{table*}
\contcaption{}
\tabcolsep=5pt
\begin{tabular}{ccrcccccccccc}
\hline
\hline
\noalign{\vspace{1pt}}
\multicolumn{6}{c}{Initial Parameters}&&\multicolumn{4}
{c}{Parameters at Period Minimum}\\
\noalign{\vspace{2pt}}
\cline{1-6}\cline{8-11}
\noalign{\vspace{2pt}}
$M_1$&$M_2$&\multicolumn{1}{c}{$P^i$}&$t^i$&$X_c^i$&$M_c^i/M$&&
$P_{\rm min}$&$M_2^{\rm min}$&$\dot{M}_{\rm min}$& $X_s^{\rm min}$&
gap&weight\\
(\Msun)&(\Msun)&\multicolumn{1}{c}{(hr)}&(yr)&&&&(min)&(\Msun)&(\Msyr)&&(hr)\\
\noalign{\vspace{2pt}}
\hline
\noalign{\vspace{2pt}}
 1.0 &  0.6 &   4.3 &  0.0E+00 &  0.685 &  0.000 &  &
 77.1 &  0.067 &  2.8E$-$11 &  0.683 &
 2.76$-$3.03 &   1.3E$-$02\\
\noalign{\vspace{5pt}}
 1.0 &  0.8 &   5.5 &  0.0E+00 &  0.685 &  0.000 &  &
 77.1 &  0.067 &  2.7E$-$11 &  0.682 &
 2.65$-$3.01 &    6.4E$-$03\\
 1.0 &  0.8 &   5.8 &  3.9E+09 &  0.560 &  0.000 &  &
 76.2 &  0.066 &  2.7E$-$11 &  0.664 &
 2.51$-$2.87 &   5.7E$-$03\\
 1.0 &  0.8 &   6.0 &  8.0E+09 &  0.436 &  0.000 &  &
 75.7 &  0.065 &  2.6E$-$11 &  0.643 &
 2.29$-$2.62 &   1.1E$-$03\\
\noalign{\vspace{5pt}}
 1.0 &  0.9 &   6.1 &  0.0E+00 &  0.685 &  0.000 &  &
 77.1 &  0.066 &  2.5E$-$11 &  0.682 &
 2.66$-$3.01 &    4.5E$-$03\\
 1.0 &  0.9 &   6.4 &  2.7E+09 &  0.552 &  0.000 &  &
 75.8 &  0.066 &  2.6E$-$11 &  0.661 &
 2.49$-$2.84 &    2.7E$-$03\\
 1.0 &  0.9 &   6.8 &  5.1E+09 &  0.433 &  0.000 &  &
 75.8 &  0.063 &  2.8E$-$11 &  0.639 &
 2.22$-$2.56 &    7.6E$-$04\\
 1.0 &  0.9 &   7.2 &  7.7E+09 &  0.310 &  0.000 &  &
 73.3 &  0.063 &  2.6E$-$11 &  0.612 &
 1.69$-$1.88 &    5.6E$-$04\\
 1.0 &  0.9 &   8.5 &  1.2E+10 &  0.093 &  0.000 &  &
 64.6 &  0.033 &  1.1E$-$11 &  0.394 &
  &     2.5E$-$04\\
\noalign{\vspace{5pt}}
 1.0 &  1.0 &   6.9 &  0.0E+00 &  0.685 &  0.000 &  &
 77.8 &  0.070 &  3.4E$-$11 &  0.682 &
 2.67$-$3.01 &  4.0E$-$03\\
 1.0 &  1.0 &   7.3 &  1.5E+09 &  0.567 &  0.000 &  &
 75.5 &  0.067 &  2.7E$-$11 &  0.662 &
 2.50$-$2.88 &    2.7E$-$03\\
 1.0 &  1.0 &   7.7 &  3.2E+09 &  0.444 &  0.000 &  &
 75.0 &  0.066 &  2.8E$-$11 &  0.638 &
 2.16$-$2.50 &   7.5E$-$04\\
 1.0 &  1.0 &   8.2 &  4.6E+09 &  0.327 &  0.000 &  &
 72.9 &  0.061 &  2.5E$-$11 &  0.595 &
 1.42$-$1.57 &    7.7E$-$04\\
 1.0 &  1.0 &   9.5 &  7.1E+09 &  0.101 &  0.000 &  &
 67.8 &  0.052 &  2.4E$-$11 &  0.483 &
 1.56$-$1.59 &    7.1E$-$04\\
 1.0 &  1.0 &  11.4 &  9.0E+09 &  0.000 &  0.000 &  &
 53.9 &  0.058 &  8.9E$-$11 &  0.316 &
  &    3.5E$-$04\\
 1.0 &  1.0 &  13.9 &  1.0E+10 &  0.000 &  0.034 &  &
 22.0 &  0.063 &  8.9E$-$10 &  0.034 &
  &    1.7E$-$04\\
 1.0 &  1.0 &  16.3 &  1.1E+10 &  0.000 &  0.060 &  &
  7.6 &  0.125 &  4.9E$-$08 &  0.000 &
  &    9.4E$-$05\\
 1.0 &  1.0 &  19.5 &  1.2E+10 &  0.000 &  0.093 &  & \\
\noalign{\vspace{5pt}}
 1.0 &  1.1 &   7.9 &  0.0E+00 &  0.685 &  0.000 &  &
 77.1 &  0.067 &  2.6E$-$11 &  0.682 &
 2.59$-$3.02 &   4.4E$-$03\\
 1.0 &  1.1 &   8.4 &  1.1E+09 &  0.558 &  0.000 &  &
 75.8 &  0.065 &  2.6E$-$11 &  0.659 &
 2.49$-$2.83 &   2.3E$-$03\\
 1.0 &  1.1 &   8.8 &  2.1E+09 &  0.448 &  0.000 &  &
 75.5 &  0.064 &  2.7E$-$11 &  0.637 &
 2.25$-$2.53 &    6.6E$-$04\\
 1.0 &  1.1 &   9.3 &  2.9E+09 &  0.340 &  0.000 &  &
 73.6 &  0.062 &  2.5E$-$11 &  0.615 &
 1.84$-$2.05 &   6.6E$-$04\\
 1.0 &  1.1 &  10.4 &  4.3E+09 &  0.105 &  0.000 &  &
 70.3 &  0.060 &  2.9E$-$11 &  0.534 &
 1.34$-$1.42 &    6.2E$-$04\\
 1.0 &  1.1 &  12.3 &  5.6E+09 &  0.000 &  0.000 &  &
 60.0 &  0.032 &  1.3E$-$11 &  0.321 &
  &     4.7E$-$04\\
 1.0 &  1.1 &  16.4 &  7.1E+09 &  0.000 &  0.038 &  &
 15.6 &  0.081 &  3.2E$-$09 &  0.032 &
  &     2.5E$-$04\\
 1.0 &  1.1 &  19.4 &  7.5E+09 &  0.000 &  0.072 &  & \\
\noalign{\vspace{5pt}}
 1.0 &  1.2 &   9.0 &  0.0E+00 &  0.685 &  0.000 &  &
 78.2 &  0.067 &  2.6E$-$11 &  0.682 &
 2.58$-$3.05 &    6.0E$-$03\\
 1.0 &  1.2 &  10.0 &  1.4E+09 &  0.567 &  0.000 &  &
 74.3 &  0.065 &  2.8E$-$11 &  0.636 &
 2.08$-$2.40 &    4.3E$-$03\\
 1.0 &  1.2 &  11.3 &  2.7E+09 &  0.447 &  0.000 &  &
 70.6 &  0.069 &  3.3E-11 &  0.562 &
 1.39$-$1.50 &    1.1E$-$03\\
 1.0 &  1.2 &  12.8 &  3.7E+09 &  0.329 &  0.000 &  &
 63.2 &  0.051 &  2.5E$-$11 &  0.417 &
 1.92$-$2.11 &    1.4E$-$03\\
 1.0 &  1.2 &  15.9 &  5.2E+09 &  0.105 &  0.000 &  &
 43.5 &  0.048 &  4.4E$-$11 &  0.152 &
  &     6.5E$-$04\\
 1.0 &  1.2 &  16.7 &  5.4E+09 &  0.047 &  0.000 &  &
 23.1 &  0.074 &  9.6E$-$10 &  0.031 &
  &     1.6E$-$04\\
 1.0 &  1.2 &  17.5 &  5.6E+09 &  0.010 &  0.000 &  & \\
\noalign{\vspace{5pt}}
 1.0 &  1.4 &  10.8 &  2.1E+07 &  0.680 &  0.000 &  &
 76.2 &  0.065 &  2.8E$-$11 &  0.637 &
 2.43$-$2.84 &    3.8E$-$03\\
 1.0 &  1.4 &  12.5 &  1.0E+09 &  0.548 &  0.000 &  &
 71.9 &  0.061 &  2.7E$-$11 &  0.557 &
 2.19$-$2.52 &   2.4E$-$03\\
 1.0 &  1.4 &  14.4 &  1.7E+09 &  0.436 &  0.000 &  &
 68.7 &  0.057 &  2.8E$-$11 &  0.500 &
 2.08$-$2.34 &    6.9E$-$04\\
 1.0 &  1.4 &  16.5 &  2.2E+09 &  0.340 &  0.000 &  &
 65.6 &  0.052 &  2.6E$-$11 &  0.436 &
 1.94$-$2.11 &    5.2E$-$04\\
 1.0 &  1.4 &  19.3 &  2.7E+09 &  0.221 &  0.000 &  &
 54.3 &  0.054 &  4.2E$-$11 &  0.255 &
 1.66$-$1.67 &    4.4E$-$04\\
 1.0 &  1.4 &  21.9 &  3.1E+09 &  0.112 &  0.000 &  &
 38.5 &  0.052 &  1.1E$-$10 &  0.090 &
  &     2.4E$-$04\\
 1.0 &  1.4 &  23.9 &  3.3E+09 &  0.023 &  0.000 &  & \\

\noalign{\vspace{2pt}}
\hline
\end{tabular}\\
\parbox{6truein}{
\noindent{Note. ---} $M_1\,$: mass of white dwarf; $M_2\,$: initial
mass of secondary; $P^i\,$: initial orbital period; $t^i$\,: age
at beginning of mass transfer; $X_c^i$\,: initial central hydrogen
mass fraction; $M_c^i/M$\,: initial fractional mass of the H-exhausted
core; $P_{\rm min}\,$: minimum period;  $M_2^{\rm min}$\,: secondary mass
at period minimum, $\dot{M}_{\rm min}$\,: mass-transfer
rate at period minimum; $X_s^{\rm min}$: surface H abundance
at period minimum; gap: period gap; $M_2^{\rm gap}$\,: secondary mass
at beginning of period gap; relative weight of sequence in model (1).}
\end{table*}
\vfill

\label{lastpage}

\end{document}